# The future of brain health: from developmental insights to clinical translation

Mariela Chertoff · Martin G. Frasch · Eduardo T. Cánepa · Gerlinde A. S. Metz · Marta C. Antonelli · Sheehan D. Fisher · Bea R. H. Van den Bergh

**Abstract**

Early-life adversity is a well-established correlate of later mental ill-health, yet developmental outcomes are highly heterogeneous. This review advances a developmental brain health framework that shifts emphasis from disorder endpoints to lifespan regulatory capacity, integrating evidence across developmental neuroscience, clinical science, translational psychiatry, public health and social context (section 1). We outline core principles of developmental plasticity, including critical and sensitive periods and the distinction between experience-expectant and experience-dependent processes (section 2). We synthesise how early adversity becomes clinically expressed and biologically embedded through interacting behavioural, epigenetic, neuroimmune, and neurosocial pathways. By distinguishing downstream neurobiological correlates of risk from mechanisms that regulate plasticity itself, we delineate where human evidence is strong and where inference relies on biologically proximal experimental models (section 3). The contribution of this review is integrative and translational rather than proposing a new theory: we clarify how emerging regulatory mechanisms, including multisystem stress regulation, social neurobiological pathways, and neuroimmune modulation, fit within established developmental frameworks and inform brain health-relevant intervention targets. Complementary developmental-adaptation models are integrated as linked levels of description, converging on a life-course perspective in which brain health emerges from cumulative, timing-sensitive biology-experience interactions (section 4). Translation is anchored in concrete developmental pathways, with a focused attention to parental mental health, pregnancy, and parent-child co-regulation as modifiable contexts through which supportive environments may recalibrate regulatory systems (section 5). We conclude with timing-sensitive, family-inclusive implications for prevention (section 6), and priorities for testable, lifespan-oriented models supported by multimodal biomarkers and equitable policy (section 7).

## Brain health as dynamic adaptation

Brain health is increasingly conceptualised not as a static endpoint but as a dynamic process of adaptation and regulation across the lifespan, reflecting the capacity to maintain functional integrity and behavioural flexibility in response to internal and external demands [1]. From this perspective, the brain can be understood as a self-organising complex system, in which coherent patterns of cognition, emotion, and behaviour emerge from ongoing interactions among distributed neural circuits, physiological regulatory systems, and environmental context, rather than from a single central controller [2].

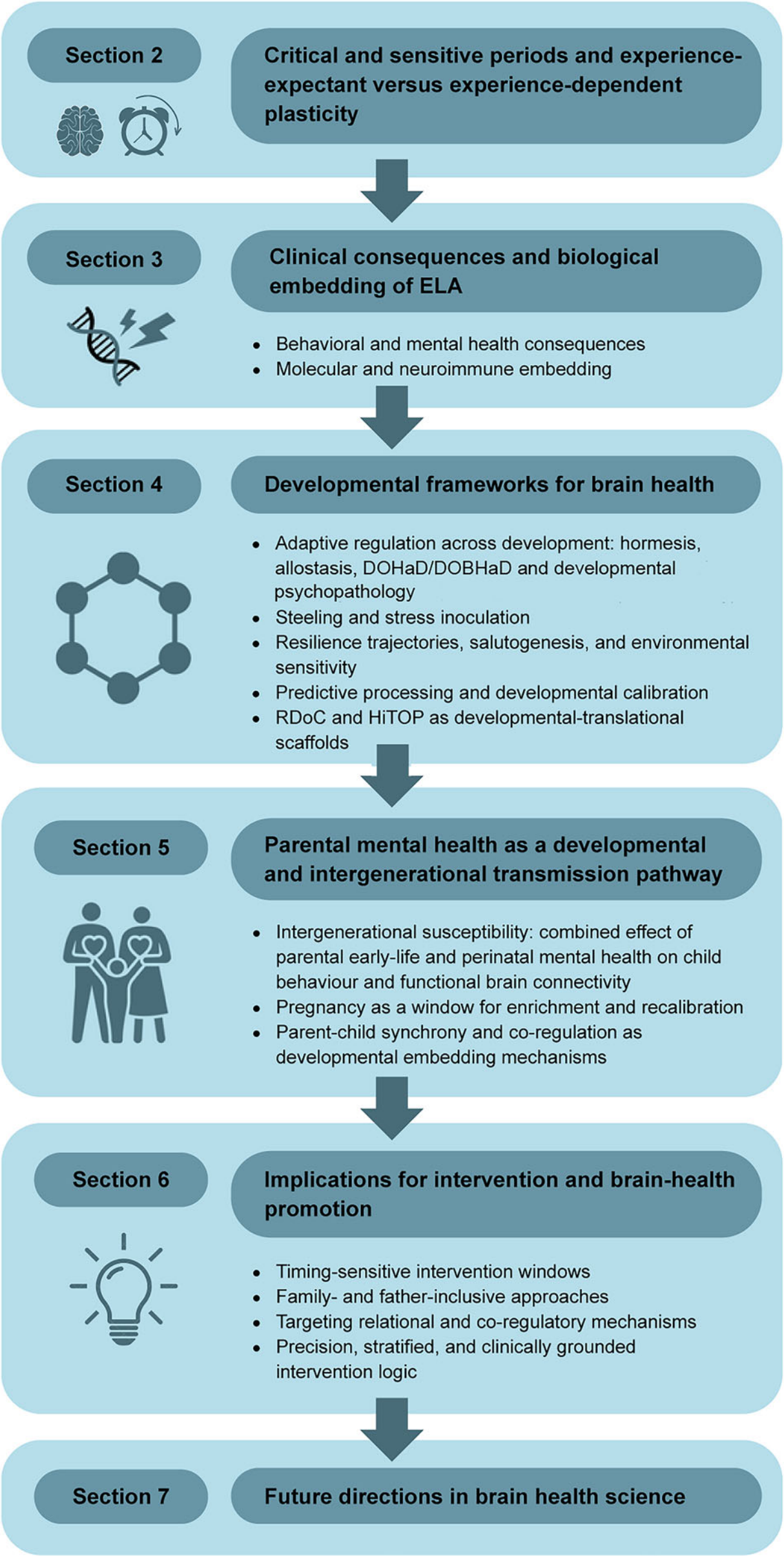

**Fig. 1** The organization and oultine of sections are depicted in Fig. 1.

Accordingly, brain health encompasses cognitive, emotional, and sensorimotor functions supported by distributed neural circuits that are continuously shaped by psychosocial, environmental, and physiological influences [3–8]. From early prenatal development onward [9], these processes unfold through reciprocal interactions between biological factors and contextual conditions, with effects that depend critically on developmental timing [10–12]. Brain health is therefore not reducible to the absence of disorder, but reflects the capacity for adaptive regulation across changing developmental demands.

The review establishes a developmental translational framework that integrates principles of developmental plasticity with contemporary models of biological embedding and clinical heterogeneity. After outlining core plasticity principles (Section 2), we examine how early-life adversity becomes expressed across behavioural, neurobiological, and multisystem pathways (Section 3), and synthesise complementary conceptual frameworks to clarify when and how early environments shape regulatory trajectories (Section 4). We then illustrate these mechanisms through concrete developmental pathways, focusing on parental mental health, pregnancy, and parent-child co-regulation (Section 5), before translating these insights into implications for intervention (Section 6) and future brain health research (Section 7). The overview is depicted in Fig. 1.

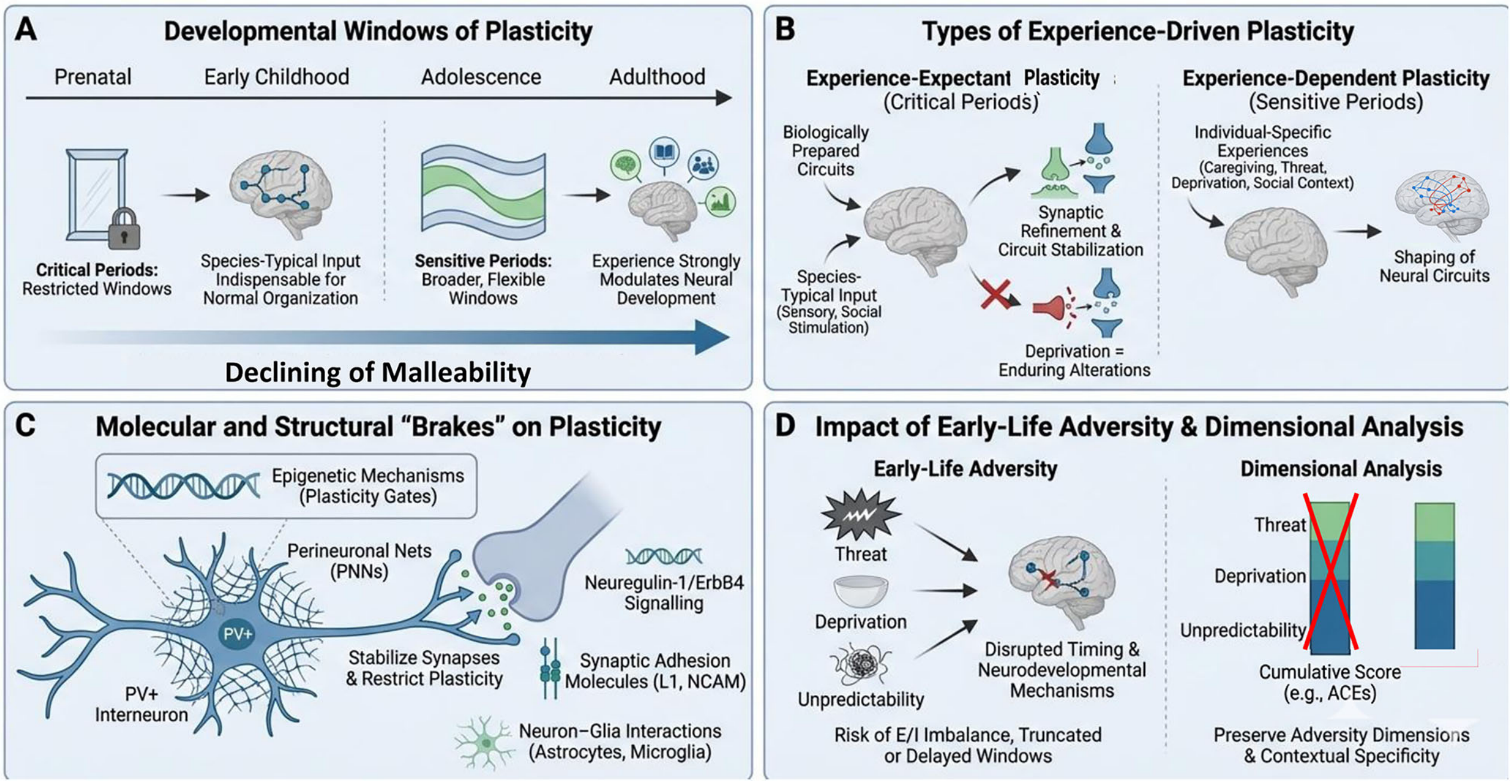

**Fig. 2 Mechanisms of Developmental Plasticity and the Biological Embedding of Experience. A** Developmental Windows of Plasticity. Neural plasticity operates along a continuum of malleability that progressively diminishes from the prenatal stage through to adulthood. A distinction are made between critical periods, with restricted windows indispensable for species-typical structural organization, and sensitive periods, which represent broader, more flexible windows where experience strongly modulates neural development. **B** Types of Experience-Driven Plasticity. Processes are categorized into: 1) Experience-expectant plasticity, where biologically prepared circuits require species-typical sensory and social input for synaptic refinement and circuit stabilization; deprivation during these phases results in enduring functional alterations. 2) Experience-dependent plasticity, where individual-specific experiences uniquely shape the architecture and connectivity of neural circuits. **C** Molecular and Structural "Brakes" on Plasticity. Schematic representation of the mechanisms that stabilize synapses and restrict further remodeling in maturity. These include epigenetic mechanisms acting as "plasticity gates", the consolidation of perineuronal nets (PNNs) around PV+ interneurons, Neuregulin-1/ErbB4 signaling, synaptic adhesion molecules (L1, NCAM), and neuro-glial interactions involving astrocytes and microglia. **D** Impact of early-life adversity and dimensional analysis. Exposure to specific dimensions of adversity: threat, deprivation, and unpredictability can disrupt the ontogenetic timing of neurodevelopmental mechanisms and alter the excitation/inhibition (E/I) balance. Dimensional analysis using a cumulative scoring system (e.g., ACEs) to evaluate how the contextual specificity of experience should not be translated into neurodevelopmental trajectories as they do not represent the influence of timing and intensity of stress.

## Terminological and conceptual distinctions

In this review, stress is defined as a process by which environmental demands, whether acute or chronic, social or biological, challenge an individual's capacity for allostatic regulation, requiring adaptive physiological, neural, and behavioural responses to maintain stability through change [13–15]. Stress exposure does not uniformly confer risk; rather, its developmental impact depends on timing, intensity, predictability, and the availability of regulatory resources [14–17]. When adaptive responses are proportionate and recoverable, stress may support resilience and plasticity; when demands are severe, prolonged, or poorly regulated, cumulative allostatic load can alter stress-regulatory systems and increase vulnerability for later brain health difficulties (see section 4.1).

Additionally, we distinguish between early-life adversity, early-life stress and maternal distress during pregnancy and early postnatal period [11, 13, 18]. *Early-life adversity* (ELA) is used as an umbrella term referring to a broad range of social, environmental, and psychological conditions that may disrupt typical development. *Early-life stress* denotes the biological stress processes- often indexed by activation of stress regulatory systems through which some adverse exposures exert effects and is frequently operationalised in experimental models. *Maternal psychological distress during pregnancy and the early postnatal period* (including the neonatal and infancy phases) encompasses anxiety, depression, or perceived stress, which may influence offspring development via maternal-placental-foetal pathways, postnatal caregiving and neurobiological pathways. These constructs are overlapping but not interchangeable, and their mechanisms of action may differ in timing, intensity, and biological proximity to neurodevelopmental processes.

The present review does not aim to provide a comprehensive account of prenatal exposure to parental psychological distress or on ELA and epigenetic mechanisms per se, as these topics have been reviewed extensively in detail elsewhere [19–22]. Instead, our focus is on conceptual integration across developmental frameworks and selected mechanistic insights that inform brain health science, while recognising that different forms of adversity may operate through partially distinct biological pathways.

These framing considerations position brain health as a timing-sensitive, context-embedded adaptive capacity, providing the foundation for the developmental frameworks introduced in Section 4 and their translational implications in subsequent sections.

## Critical and sensitive periods and experience-expectant versus experience-dependent plasticity

Brain development unfolds through temporally structured windows of heightened plasticity during which experience has a disproportionate influence on neural circuit organisation. Critical periods are temporally restricted developmental windows during which species-typical environmental input is indispensable for typical circuit formation, such that deprivation leads to enduring alterations in structure and function [23–26]. Sensitive periods, in contrast, represent broader and more flexible windows during which experience strongly modulates -rather than strictly determines- neural development, allowing for adaptation to individual and contextual variation (Fig. 2 A).

These temporal windows map onto two complementary forms of experience-driven plasticity: experience-expectant and experience-driven plasticity (Fig. 2B). Alongside experience independent processes (e.g., [27]), these forms guide early brain development. Experience-expectant plasticity, which dominates during critical periods, relies on biologically prepared circuits receiving species-typical input (e.g., patterned sensory stimulation or basic social interaction) and guides synaptic refinement and circuit stabilisation. In contrast, experience-dependent plasticity, which predominates during sensitive periods, reflects the shaping of neural circuits by individual-specific experiences, including variation in caregiving quality, threat exposure, deprivation, unpredictability, and broader social context [24, 28, 29]. Together, these processes form a continuum of developmental plasticity that extends into adulthood, but is most pronounced -and most susceptible to environmental influences, whether adverse or supportive- during early development and adolescence [30]. In adolescence, ongoing synaptic pruning, progressive myelination, and rebalancing of excitation/inhibition dynamics, particularly within association and prefrontal cortices, support executive control, emotion regulation, and goal directed behaviour, rendering this phase both a period of increased vulnerability and a potential window for recalibration following earlier perturbations [31, 32].

Importantly, for an experience to be experience-expectant, the developing brain must be biologically prepared to receive it as a typical developmental signal- a criterion that many forms of adversity do not meet. This supports the view that most adverse exposures act as experience-dependent modulators of plasticity or predictive calibration rather than as expected inputs shaping circuit organisation through canonical critical-period mechanisms ([30] and see Section 4.4).

The opening and closure of critical and sensitive periods are regulated by region-specific molecular and structural mechanisms that actively gate plasticity (Fig. 2C). Key regulators include the maturation of inhibitory circuitry, particularly parvalbumin-positive (PV + ) interneurons, and the formation of perineuronal nets (PNNs), which stabilise synaptic contacts and restrict further plasticity [33–36]. Notably, the same PV+ interneurons that initiate

**Table 1** Mental health interventions for cognitive and emotional disorders originating in early life.

| Population/Target | Area / Topic | Main Objective | Type of Study | Refs |
|---|---|---|---|---|
| Pregnant women with depression/anxiety | Prenatal exercise | Reduce prenatal symptoms | Supervised exercise | [209] |
| Foetuses and newborns | Prenatal stress and foetal programming | To investigate how prenatal stress and foetal signalling influence future mental health vulnerability | Experimental / biomarker-based | [210–212] |
| Early childhood (infants to school age) | Prevention of mental disorders in early childhood at population level | Argue for primary and secondary prevention targeting brain development, self-regulation, anti-inflammatory/neuroprotective strategies, and school-based programs | Perspective / prevention roadmap | [213] |
| Children 0–3 years and caregivers (parents or caregivers in health systems) | Early-childhood prevention 0–3 years | Summarize perinatal and early-parenting programs (attachment, reflective functioning, co-parenting) that prevent later mental health problems and support socio-emotional development | Narrative review of intervention trials and programs | [214] |
| | Multi-component parenting + parental mental health interventions | Evaluate interventions that jointly target parenting behaviour and caregiver mental health on child cognitive and socio-emotional outcomes | Systematic review & meta-analysis of RCTs | [215] |
| | Early childhood development interventions by healthcare providers | Test whether provider-delivered responsive caregiving/early learning interventions improve cognitive and motor development and maternal mental health | Systematic review & meta-analysis of RCTs | [216] |
| | Cortisol and parenting style | Improve stress regulation | Parenting style modification based on cortisol reactivity | [217] |
| Children with emotional/behavioural dysregulation | Real-time wearable technology augmentation of parent-child intervention | Test feasibility/effectiveness of a smartwatch-alert system to shorten tantrums and improve parent response within Parent-Child Interaction Therapy (PCIT). | Randomized clinical trial | [218] |
| | Mental health, earlier – self regulation in primary care | Early identification (neurodevelopmental risk markers) and brief self- regulation interventions to prevent later emotional/behavioral disorders | Conceptual translational roadmap, synthesis of trials | [219] |
| | Early precursors & interventions for ADHD (0–5 yrs) | Identify early neurocognitive/behavioral precursors (executive function, regulation) and test early interventions that reduce ADHD symptoms and improve working memory | Dual systematic reviews & meta-analyses | [220] |
| | Interventions for young children's mental health | Synthesize evidence on behavioural / Cognitive behavioral therapy (CBT) based child and parent programs improving broad mental health, internalizing, externalizing, ADHD, anxiety | Umbrella "review of reviews" (systematic + meta-analytic reviews) | [221] |
| | Autism and neuro-immunometabolic origins | To present a prenatal origin model based on neuroimmune and metabolic interactions | Integrative hypothesis | [222] |
| Children/teens with mental health disorders | Universal/selective youth mental health promotion | Assess universal/selective interventions that enhance positive mental health outcomes (emotions, cognitive skills, QoL) and prevent later disorders | Large systematic review & meta-analysis of preventive programs | [223] |
| | Psychotherapy | Examine RCT outcomes on psychotherapy (CBT, DBT) for emotional/behavioral difficulties, including BPD, PTSD, conduct disorder, and suicidality. | Systematic review | [224] |
| | Diagnostic biomarkers | Improve diagnostic accuracy | Use of biomarkers for early diagnosis | [225–227] |
| Children and parents | RSA and physiological synchrony | Improve emotional regulation and bonding | Intervention based on RSA and dyadic RSA | [194, 228, 229] |
| | Parenting style and stress | Promote early child development | Parent training and support | [164, 230–237] |
| | Parent-child synchrony | Enhance bonding and emotional regulation | Training to improve behavioural and emotional synchrony | [238, 239] |
| | Fatherhood and co-parenting | Optimize paternal role and parental cooperation | Parenting and co-parenting programs | [240] |
| | Intergenerational transmission of adversity | To examine how maternal adverse childhood experiences affect child development through psychosocial pathways | Longitudinal observational | [212, 241–243] |
| Infants and young children exposed to prenatal stress/ELA | Biomarkers in prenatal intervention | Monitoring and use of epigenetic mechanisms, inflammatory markers and Heart rate variability | Identify and guide early interventions | [244, 245] |
| | ELA and Mental health | Quantify associations between childhood trauma, depression symptoms, and cognitive performance in drug-naïve adults; highlights the impact of early adversity on later mental/cognitive health. | Observational case-control study | [246] |
| | Environmental conditions for healthy brain development | Define nurturing, responsive caregiving and stimulation as modifiable environmental inputs for optimal brain/emotional development; propose routine care preventive interventions during sensitive periods | Integrative neuroscience review with intervention framework | [247] |
| | Socioeconomic disadvantage | Programs targeting poverty | Mitigate negative effects of socioeconomic context | [248] |
| | Behavioural interventions | Map types, delivery, and effectiveness of behavioural and parenting programs for early mental health problems (disruptive, anxiety, social–behavioural) | Scoping review of experimental studies | [249] |
| Adolescents and adults exposed to ELA | Pubertal recalibration | Neuroendocrine and socioemotional recalibration | Intervention during critical pubertal period (adoption) | [59, 250, 251] |

**Table 1** continued

| Population/Target | Area / Topic | Main Objective | Type of Study | Refs |
|---|---|---|---|---|
| | Emotional circuit development and unpredictability | To study amygdala activation and its link to early-life unpredictability | Functional neuroimaging | [252–254] |
| | Neurodevelopment and early adversity | To assess the impact of early adversity on brain architecture and risk for mental health disorders | Longitudinal cohort | [255–258] |
| | Neurodevelopmental optimization after early adversity | Define sensitive periods and brain-circuit targets; propose randomized enhancement trials in early childhood to optimize cognitive/emotional outcomes after adversity | Cross-species neuroscience review with intervention designs | [153, 247, 259, 260] |
| | Sex-specific and minorities intervention | Improve response,personalized treatment | Pharmacological treatments, psychosocial and therapeutic approach | [261–263] |
| Humans (clinical and theoretical) | Stress, uncertainty, and disease mechanisms | To explain how the brain manages uncertainty and stress, and how this relates to mental illness | Applied theoretical model | [264] |
| Adolescents and young adults (12–25) | Digital mental health interventions for emerging youth | Map types, delivery formats, and outcomes across digital interventions for youth mental health (prevention and treatment) to identify gaps and evidence. | Scoping review | [265] |
| | Youth mental health paradigm & early intervention | Link brain developmental windows with staging models and early, transdiagnostic, often digital interventions to prevent progression to severe disorders | Perspective / roadmap with evidence synthesis | [266–268] |
| Various populations | Multisystemic interventions | Enhance resilience and adaptive development | Integrated approach (biological-psychosocial) | [198, 236, 269–271] |
| Animal models / adult humans | Microglial memory and early-life inflammation | To explore how early-life stress induces microglial memory and increases susceptibility to neurodegeneration | Experimental / integrative | [272, 273] |
| Juvenil and adolescent *Rhesus macaques* | Stress modulation | Neuroendocrine response to dexamethasone | Experimental | [274] |

heightened plasticity later become primary targets of PNN encapsulation, illustrating a coordinated programme linking plasticity induction to its closure [34–36]. Epigenetic processes also act as plasticity gates: experimental manipulation of histone acetylation can partially reinstate juvenile-like plasticity in adult cortex, underscoring the dynamic regulation of these windows [25, 37].

Additional modulators fine-tune the timing and duration of plasticity windows. Neuregulin-1/ErbB4 signalling regulates excitatory input onto PV+ interneurons, influencing both the onset and closure of sensitive periods [38, 39]. Synaptic adhesion molecules (e.g., L1, NCAM) and neuron-glia interactions -particularly astrocytic and microglial regulation of synaptic maturation- further shape circuit refinement and stabilisation [40–42]. Disruption of these processes can have lasting consequences: delayed or insufficient excitation-inhibition (E/I) balance may prevent timely opening of plasticity windows, whereas premature stabilisation may truncate them before adequate circuit refinement occurs [33, 43–45].

Beyond psychosocial inputs, experimental studies demonstrate that biologically proximal prenatal and early postnatal exposures -such as protein-energy malnutrition- can directly alter neurodevelopmental processes that regulate sensitive and critical periods. Rodent studies show that perinatal malnutrition induces long-lasting changes in white-matter pathways, oligodendrocyte morphology, and miRNA-mediated regulation of axon guidance, with behavioural consequences that can be partially reversed by postnatal environmental enrichment [46]. Such findings align with recent integrative frameworks proposing that some prenatal environments may shape postnatal plasticity by modulating neurobiological gateways, including excitation-inhibition balance, myelination dynamics, synaptic pruning, and the molecular triggers and brakes governing sensitive-period timing [16, 30, 47, 48]. Here, we emphasise that these biologically proximal exposures probably differ in important ways from prenatal psychological distress; whether -and to what extent- some underlying mechanisms may overlap remains to be established (Fig. 2D).

Research on brain plasticity and developmental adaptation is informing novel interventions approaches (Table 1), alongside strategies for early identification and support. The table is illustrative rather than exhaustive, highlighting selected recent and relevant work.

In sum, neuroplasticity represents both a window for adaptive change and a point of heightened sensitivity; Section 3 therefore examines how ELA becomes clinically expressed and biologically embedded across behavioural, epigenetic, and neuroimmune pathways.

## Clinical consequences and biological embedding of ELA

ELA is associated with increased risk for a broad range of mental health outcomes across the lifespan, including mood and anxiety disorders, stress-related psychopathology, and impairments in socio-emotional and cognitive functioning [29, 49, 50]. Importantly, these associations do not reflect uniform deficit outcomes but rather heterogeneous developmental trajectories, underscoring the importance of developmental timing, adversity dimensions, and contextual factors in shaping risk and resilience.

ELA is often operationalised using cumulative indices such as Adverse Childhood Experiences (ACEs), which show robust associations with later mental and physical health outcomes [49]. However, aggregating heterogeneous exposures into a single score can obscure critical differences in timing, severity, chronicity, and qualitative dimensions of adversity (e.g., threat, deprivation, unpredictability), and may conflate exposure effects with correlated contextual resources and constraints [29, 51]. Developmentally informed models therefore emphasize analytic approaches that preserve adversity dimensions and developmental timing, rather than relying solely on additive exposure counts.

### Behavioural and mental health consequences

Clinical expression of ELA varies systematically by developmental timing, adversity type, and sex. Early studies suggested that prenatal ELA exposures are more consistently associated with alterations in stress regulation and hippocampal development [14, 16]. More recent longitudinal and whole brain neuroimaging studies indicate that not only prenatal exposure to ELA but also to maternal psychological distress is associated with disrupted connectivity of neural networks already in the foetus [52, 53]. Next to changes in the amygdala-frontal tract [54], the latter exposure may be associated with altered white-matter microstructure involving projection, association, and callosal tracts rather than limbic–prefrontal regions alone [55, 56]. Evidence for persistence until adulthood, however, currently derives mainly from relatively small exploratory samples and should be interpreted as preliminary [56].

Adversity during childhood and adolescence implicates amygdala-prefrontal circuitry supporting emotion regulation, threat processing, and executive control [14, 16]. Adolescence constitutes a sensitive phase in which recalibration of threat, reward, and cognitive control systems may heighten vulnerability to psychopathology, while also permitting recovery and adaptive reorganisation under supportive conditions [31, 57, 58].

Sex-specific patterns further complicate clinical prediction. Females more frequently report complex adversity profiles and show higher prevalence of affective disorders, whereas males may exhibit greater cognitive or behavioural impacts following deprivation or institutional rearing [59, 60]. These differences likely reflect interactions between developmental timing, neuroendocrine maturation, and social context.

Socioeconomic disadvantage is particularly consequential for brain health because it aggregates multiple co-occurring exposures, including material deprivation, chronic stress, environmental hazards, reduced access to enrichment, and constrained institutional support [61, 62]. Experimental models approximating this complexity demonstrate that combined social and material deprivation during gestation and early postnatal life can induce enduring alterations relevant to mental health, including reduced maternal care, sex-specific increases in anxiety-like behaviour and aggression, prefrontal and hippocampal structure and immune and chromatin-related gene regulation [63].

Consistent with this multisystem perspective, comparative WEIRD (Western, Educated, Industrialized, Rich and Democratic; [64]) versus non-WEIRD evidence indicates both shared mechanisms and context-dependent variation in the impact of perinatal maternal psychological distress on offspring outcomes (e.g., preterm birth, birth weight, neurodevelopment), shaped by socioeconomic conditions that influence exposure profiles, prevalence, and access to healthcare [65–67]. Large-scale longitudinal and neuroimaging studies similarly link socioeconomic disadvantage to differences in brain structure, functional connectivity, educational outcomes, and psychopathology risk, while highlighting substantial heterogeneity and the need to distinguish household level from neighbourhood level effects [68–70]. Crucially, increased vulnerability should not be conflated with deterministic deficit as adaptive functioning remains possible given adequate contextual supports [71].

## Molecular and neuroimmune embedding

Beyond behavioural outcomes, ELA becomes biologically embedded through coordinated changes in epigenetic regulation, neuroimmune signalling, and neuromodulatory systems. Experimental work has long demonstrated timing-dependent, region-specific alterations in synaptic and circuit maturation within prefronto-limbic networks following early-life stress and glucocorticoid exposure [72–78]. Recent integrative reviews reinforce these findings while also highlighting unresolved questions regarding stress perception, sensitive developmental windows, and translational relevance [79, 80].

Altered DNA methylation (DNAm) and microRNA (miRNA) profiles have been observed following adversity exposure, particularly in genes regulating glucocorticoid signalling (e.g., *NR3C1*, *FKBP5*), neuroplasticity, inflammation, and serotonergic function [81–84]. These epigenetic signatures may persist over long periods and are increasingly detectable in peripheral tissues and circulating exosomes [22, 85]. Multidimensional early-life stress has been shown, in mice (postnatal day 12) to induce a precocious shift in visual cortical critical-period timing, accompanied by premature formation of inhibitory perineuronal nets and metabolomic and transcriptomic signatures indicative of accelerated inhibitory stabilization [45]. However, such findings are best interpreted as downstream correlates of biological embedding rather than direct regulators of plasticity mechanisms.

Neuroimmune pathways represent a complementary mechanism of biological embedding. Chronic or severe stress exposure, particularly during sensitive developmental windows, is associated with sustained elevations in pro-inflammatory cytokines (e.g., IL-1β, IL-6), contributing to glial activation, synaptic pruning abnormalities, and impaired neurogenesis [86]. Elevated inflammatory markers have been linked to structural and functional alterations in the hippocampus and prefrontal cortex and are associated with increased risk for depression, cognitive decline, and accelerated biological aging ("inflammaging") [87, 88].

Beyond stress and immune pathways, neurosocial plasticity has emerged as a candidate mechanism linking social experience to resilience. This framework posits that the "social brain" is continuously shaped by dynamic interpersonal interactions, attachment, and social stress [89, 90]. The quality of these environments, ranging from supportive relationships to toxic experiences, modulates brain development, behavioural outcomes and healthy aging [91–93]. At the neurobiological level, interactions between brain-derived neurotrophic factor (BDNF) and oxytocin modulate synaptic plasticity, social bonding, and stress buffering across development, with age- and sex-dependent effects [2, 94].

Finally, although most ELA research focuses on downstream developmental outcomes, a smaller but growing literature suggests that certain early environments may also influence the neurobiological mechanisms that regulate plasticity itself. This inference derives largely from experimental models of biologically proximal exposures, rather than psychosocial adversity, and evidence in humans remains limited and exposure-specific [16, 30, 47](see also section 2). In particular, there is sparse support for comparable effects of prenatal exposure to maternal psychological distress [30, 95]. This gap highlights the need to examine how different forms and timings of adversity interact with plasticity-regulating systems across development.

In summary, ELA gives rise to diverse clinical trajectories and becomes biologically embedded through interacting behavioural, epigenetic, neuroimmune, and neurosocial pathways rather than a single mechanism. Section 4 therefore turns to renewed conceptual frameworks that aim to clarify when early adversity constrains regulatory flexibility versus when adaptive calibration is preserved and how these insights can be translated clinically.

## Developmental frameworks for brain health

A growing body of developmental, clinical, and translational research indicates that early-life stress and ELA (see definitions in section 1.1) do not exert uniform or inevitably negative effects on brain health. Instead, their impact depends on developmental timing, dose, and systemic context, particularly during periods of heightened plasticity [28, 30]. These insights have motivated frameworks that move beyond cumulative-risk or deficit models toward adaptive regulation, flexibility, and resilience. The frameworks reviewed below represent complementary lenses operating at different explanatory levels rather than competing accounts.

### Adaptive regulation across development: hormesis, allostasis, DOHaD/DOBHaD and developmental psychopathology

*Hormesis* refers to biphasic dose-response relationships in which low or intermittent challenges may elicit adaptive cellular or physiological responses, whereas higher, chronic or poorly regulated loads become maladaptive or toxic [96–98]. In brain health research, such hormetic effects are most robustly demonstrated for biologically controllable challenges -including physical exercise, metabolic stressors, and environmental enrichment- which engage neurotrophic, mitochondrial, and immunometabolic pathways supporting neural reserve and functional flexibility [98–102]. In contrast to biologically controllable challenges, psychosocial adversity and maternal psychological distress constitute multidimensional exposures that are not precisely dosable, controllable or isolated, and therefore cannot be straightforwardly mapped onto classical hermetic models.

Developmental adaptation to such exposures is more appropriately conceptualised through *allostasis*, defined as the active process by which organisms achieve stability by dynamically adjusting physiological and behavioural systems across time [13]. Within this framework, allostatic load refers to the cumulative physiological costs incurred when stress-regulatory adaptations are frequent, prolonged, or poorly regulated [13]. From an allostatic perspective, one major risk associated with early adversity lies in cumulative physiological "wear and tear", which may constrain regulatory flexibility and plasticity across development. Accordingly, McEwen and colleagues, emphasised that chronic, unpredictable, or uncontrollable adversity, including sustained maternal psychological distress, may increase allostatic load and overload [17]. Rutter's work further cautions against deterministic interpretations by conceptualising resilience as a dynamic process, in which adversity may be associated with divergent outcomes depending on individual, contextual, and developmental conditions [103]. Neither perspective explicitly frames these processes in hormetic terms. Conceptually, however, their emphasis on sustained or excessive adversity, regulatory burden, and maladaptive outcomes aligns more closely with the toxic or excess arm of dose-response models, than with the low-dose adaptive responses described in hormetic frameworks. Where hormetic-like effects are observed, they should therefore be understood as conditional and system-specific, embedded within broader multisystem regulation rather than as universal explanations for ELA [102].

The *Developmental Origins of Health and Disease* (DOHaD) framework provides a unifying account of how early environments become biologically embedded, originally emphasising how prenatal conditions shape foetal physiology and regulatory systems with lasting consequences for health across the lifespan [104]. Central to DOHaD, including the concept of predictive adaptive responses, is the idea that developmental adaptations are probabilistic and context-dependent, such that early calibration of regulatory systems may confer short-term functional fit while also entailing long-term trade-offs, particularly when postnatal environments diverge from prenatal cues. Extensions such as the *Developmental Origins of Behaviour, Health, and Disease* (DOBHaD) framework explicitly place behaviour and the developing brain and (epi)genome at the centre of these processes, highlighting pregnancy and early postnatal life as periods of both heightened opportunity and vulnerability [105]. Within this perspective, maternal psychological distress is conceptualised not as a direct biological stressor but as a contextual psychological exposure that may bias foetal regulatory calibration and interact with postnatal experience-dependent plasticity rather than determining fixed outcomes [47, 106].

These biological programming and regulatory models converge with the *developmental psychopathology* framework, which conceptualises mental health and disorder as emerging from dynamic, timing-sensitive interactions between the developing organism and its environment [107, 108]. Core principles such as equifinality and multifinality explain why similar early conditions can lead to divergent developmental trajectories depending on timing, context, and regulatory capacity [109, 110]. Although prenatal influences were incorporated relatively late into mainstream developmental psychopathology, experimental and observational work has long demonstrated that organism-environment coactions operate already in utero, including effects of maternal emotional states on foetal behaviour and regulation [111–118]. Together, DOHaD/DOBHaD and developmental psychopathology provide a complementary supra-synthesis linking early biological embedding with postnatal developmental trajectories.

From an evolutionary-developmental perspective, Del Giudice's adaptive calibration model complements these frameworks by specifying how early environmental cues may tune stress-regulatory, life-history, and socio-emotional strategies to anticipated ecological demands [119, 120]. This view aligns with DOHaD/DOBHaD, allostatic perspective and predictive processing (see Section 4.4) accounts insofar as adversity -including prenatal maternal distress- is conceptualised as an experience-dependent cue shaping regulatory calibration, rather than as an experience-expectant developmental input [24, 25, 30].

Taken together these frameworks converge on a life-course health development perspective, in which brain health emerges through cumulative, timing-sensitive interactions between biology, experience, and context across preconception, prenatal, and postnatal periods, rather than from the effects of isolated exposures [121, 122].

## Steeling and stress inoculation

Within resilience research, hormesis-related ideas partly overlap with steeling effects and stress inoculation models, which propose that manageable stressors may, under specific conditions, enhance later coping, whereas overwhelming exposures lead to sensitisation [123, 124]. Population-level reports of non-linear adversity-outcome associations have sometimes been interpreted in this light [125], but such findings require cautious interpretation due to selection effects and unmeasured contextual resources.

From a developmental perspective, beneficial challenge is not equivalent to low adversity or simple recovery after stress. Where steeling-compatible effects are observed, they are most plausibly linked to calibration or training processes -including changes in regulatory set points, coping repertoires, and physiological flexibility- whose expression depends on developmental timing, controllability, and multisystem support [126, 127].

## Resilience trajectories, salutogenesis, and environmental sensitivity

Resilience-oriented and salutogenic frameworks emphasise that adaptive functioning emerges from dynamic interactions among neural systems, caregiving relationships, social environments, and broader structural conditions, rather than from stress exposure alone [127–130]. From this perspective, resilience reflects preserved or restored regulatory capacity and flexible engagement of cognitive, emotional, and social resources over time.

Longitudinal research consistently identifies heterogeneous trajectories following early adversity -including minimal impact resilience, recovery, delayed disruption, and chronic impairment- supporting resilience as a dynamic course of functioning rather than a fixed trait [131, 132]. This trajectory-based perspective distinguishes resilience from hormetic or steeling accounts by focusing on regulatory flexibility and recovery processes rather than exposure-dependent strengthening. Recent syntheses further characterise resilience as an active, multisystem process supported by interacting neural, immune, metabolic, and gut-brain pathways, with implications for biologically informed prevention [133].

Salutogenic frameworks extend this view by shifting attention from risk prediction to the generation and maintenance of health-promoting resources. Central to salutogenesis is the sense of coherence -the extent to which experiences are perceived as comprehensible, manageable, and meaningful- which is associated with more effective stress regulation and long-term wellbeing [134, 135].

Frameworks of differential environmental sensitivity further refine this picture by emphasising individual differences in responsiveness to environmental inputs, for better and for worse [136, 137]. These models complement allostatic and hormetic accounts by specifying *for whom* environments matter most, rather than *how* stressors regulate biological systems.

## Predictive processing and developmental calibration

Predictive processing provides a computational framework for understanding adaptation across development, conceptualising the brain as a hierarchical prediction system that minimises prediction error through precision-weighted updating of internal models [138, 139]. Within affective and interoceptive domains, this framework emphasises regulation through the weighting and interpretation of incoming signals relative to prior expectations [140, 141].

Recent theoretical and empirical work highlights that interoception depends not on signal accuracy alone, but on flexible adaptation of interoceptive precision across contexts and timescales [142]. Consistent with this view, transdiagnostic studies show impaired updating of interoceptive precision across affective and somatic symptom domains and experimental work in health anxiety demonstrates that imprecise expectations about bodily sensations amplify symptom-related uncertainty [143].

From a developmental perspective, predictive processing offers a formal language for linking experience-dependent learning, stress sensitisation, and regulatory rigidity, while avoiding the assumption that adversity constitutes an experience-expectant developmental input (a distinction emphasised in models of sensitive periods [30]; see section 2). At the same time, predictive processing remains a broad framework whose translational value depends on explicit model specification, falsifiable hypotheses, and clear mappings to clinically meaningful outcomes [144–146].

### RDoC and HiTOP as developmental-translational scaffolds

Traditional diagnostic systems such as the Diagnostic and Statistical Manual of Mental Disorders (DSM) and International Classification of Diseases (ICD) have been criticised for categorical, symptom-based classifications that insufficiently capture developmental change, transdiagnostic mechanisms, and underlying neurobiological processes [147, 148]. These limitations have motivated dimensional and mechanism-oriented frameworks such as the Research Domain Criteria (RDoC) and the Hierarchical Taxonomy of Psychopathology (HiTOP) [147–149].

RDoC organises core psychological and neurobiological functions -such as threat processing, reward learning, cognitive control, and social processes- across multiple units of analysis, explicitly framing these domains as developmentally embedded and experience-sensitive [150–152]. Integrative work further links RDoC dimensions to environmental features such as threat, deprivation, unpredictability, and positive input, highlighting trade-offs between short-term adaptation and long-term brain health risk [153–155].

In contrast, HiTOP provides a statistically grounded organisation of symptom covariance and clinical course [156], and shared genetic liability. Although partially convergent, the two frameworks diverge in emphasis, with RDoC constructs mapping more directly onto biological and developmental mechanisms [157–159].

Together, RDoC and HiTOP provide complementary development translational scaffolds: HiTOP is particularly informative for describing phenotypic organisation, whereas RDoC is better suited for interrogating developmental mechanisms and identifying intervention targets. Their integration supports advances in prevention, precision psychiatry, and brain health promotion across the lifespan [160–162].

Rather than advancing a single explanatory theory, this section converges on a common translational implication: developmental trajectories are shaped through modifiable relational and regulatory contexts, early intervention should target not only the developing child directly, but also the relational and regulatory contexts through which child plasticity are shaped. Section 5 therefore turns to parental mental health, pregnancy, and parent-child co-regulation, which are highlighted not because they exhaust possible intervention targets, but because they are modifiable timing-sensitive pathways through which stress regulation, social learning, and biological embedding unfold across generations.

## Parental mental health as a developmental and intergenerational transmission pathway

### Intergenerational susceptibility: combined effect of parental early-life and perinatal mental health on child behaviour and functional brain connectivity

As outlined in Section 4, adaptive versus maladaptive stress calibration depends on prior developmental history, timing, and context. From a life-course health development perspective, parental mental health during the perinatal period reflects not only concurrent stressors but also cumulative regulatory shaping across earlier life stages, including preconception and interconception health [121, 163]. Accordingly, beyond established risk factors such as a history of mood or anxiety disorders, postpartum depression, or heightened sensitivity to hormonal fluctuations (e.g., premenstrual dysphoric disorder) [164–166], early adversity appears to create cognitive and affective vulnerabilities that interact with perinatal stressors. For example, maternal early adversity has been linked to postpartum depressive symptoms via mechanisms such as repetitive negative thinking, indicating biased stress processing rather than hormetic "steeling" [167]. Consistent with this view, maternal childhood maltreatment predicts maladaptive infant socioemotional outcomes (with household dysfunction exerting indirect effects via maternal mental health [168]) as well as internalizing problems at 8 years [169]. Moreover, Uy et al. [170] showed that poorer maternal postnatal mental health was uniquely associated with more negative child fronto-amygdala resting-state functional connectivity and emotional outcomes, independent of maternal childhood trauma and prenatal mental health [169].

This cumulative, timing-sensitive pattern is further supported by intergenerational evidence showing that caregiving practices are transmitted across generations: grandparental caregiving is associated with fewer internalising

### Box 1: Factors influencing brain development and behaviour

- Epigenetic regulation: [21, 22, 85, 177, 275–282]
- Neuroimmune signalling: [283–288]
- Foetal nutrient/hormone exposure, placental endocrine-immune interaction: [19, 21, 91–93, 167]
- Neurotrophic and oxidative stress pathways: [94, 289–298]

and externalising problems in grandchildren, both directly and indirectly through continuity in parenting behaviours across generations [171]. Such findings underscore that susceptibility and resilience are shaped across linked developmental periods rather than arising from isolated perinatal exposures.

Although research has historically focused on maternal pathways, growing evidence highlights paternal contributions to intergenerational risk transmission. Paternal early adversity has been associated with alterations in neural mechanisms underlying facial emotion recognition [172], and a recent meta-analysis demonstrated that paternal perinatal distress is independently associated with poorer socio-emotional, cognitive, and global developmental outcomes in offspring, particularly when distress persists into the postpartum period [173]. While positive paternal involvement is linked to favourable developmental outcomes, perinatal paternal mental illness may reduce engagement and compromise caregiving quality [174]. Together, these findings underscore that intergenerational susceptibility is best understood within a life-course family-level framework, in which parental mental health reflects cumulative developmental influences and shapes offspring brain-behaviour trajectories through both maternal and paternal pathways, consistent with multisystem resilience perspectives (Section 4.3).

## Pregnancy as a window for enrichment and recalibration

Pregnancy represents a developmentally sensitive period in which earlier regulatory histories intersect with current psychosocial context to influence maternal-foetal processes, consistent with experience-dependent plasticity described in Section 2. Positive emotional, social, and sensory experiences during pregnancy have been associated with improved neuroendocrine balance, reduced stress hormone exposure, and more optimal placental functioning [175].

Favourable maternal emotional states during pregnancy have been linked to longer gestational duration, higher birth weight, and reduced risk of preterm birth [134, 176]. These associations likely reflect interacting pathways involving stress physiology, immune modulation, metabolic regulation, and epigenetic mechanisms rather than simple exposure-outcome effects [135, 177, 178] (see also Box 1). Longitudinal evidence further suggests that higher perceived social support and positive maternal mental health during pregnancy are associated with better cognitive performance and reduced risk of mental and behavioural disorders in children, even in the presence of maternal depressive symptoms [179, 180]. Although genetic and environmental influences cannot be fully disentangled, these findings reinforce pregnancy as a window of opportunity in which supportive contexts may recalibrate stress regulatory systems with lasting implications for offspring brain health.

## Parent-child synchrony and co-regulation as developmental embedding mechanisms

Evidence across diverse family contexts, including adolescent mothers, fathers, and structurally constrained caregiving environments, indicates that parenting practices, co-regulation, and relational support during the peripartum and early postnatal period are central modifiers of child stress regulation and mental health risk across both WEIRD and non-WEIRD samples [170, 181–184].

Biobehavioural synchrony and co-regulation constitute central postnatal mechanisms through which parental mental health and early environments become biologically embedded in the developing child. Through repeated, temporally coordinated exchanges of behaviour, physiology, and affect, caregivers scaffold infants' still-immature regulatory systems, shaping stress responsivity, emotion regulation, and the development of social brain networks [185, 186]. From a developmental brain health perspective, synchrony represents a concrete instantiation of experience-dependent plasticity operating during sensitive periods, translating relational input into enduring regulatory capacities.

Although synchrony has been most intensively studied in childhood, it is increasingly conceptualised as a developmentally continuous process that is reorganised -rather than replaced- across childhood and adolescence, in parallel with changes in neurobiological and socio-cognitive demands. Early synchrony in autonomic functioning, particularly indices of parasympathetic regulation such as respiratory sinus arrhythmia (RSA), has been associated with emerging emotion regulation capacities, stress flexibility, and later adaptive functioning [166, 187]. At the neural level, synchronous caregiving interactions are thought to support the organisation and calibration of prefrontal- limbic circuits involved in emotion regulation, reward processing, and social cognition, with oxytocinergic and vagal pathways facilitating plasticity during sensitive developmental windows [186, 188].

Longitudinal neuroimaging evidence provides direct support for these mechanisms. Using functional near-infrared spectroscopy, Stern et al. [189] showed that maternal sensitivity during early childhood was associated with stronger and increasing dorsolateral prefrontal cortex responses to positive emotional stimuli between 5 and 7 months of age. These findings suggest that sensitive caregiving may actively scaffold early prefrontal-emotional calibration, shaping receptivity to positive social cues during a sensitive period and underscoring that developmental susceptibility entails openness to supportive input as well as vulnerability to adversity (see Sections 2 and 4).

Maternal and paternal contributions to synchrony appear complementary rather than redundant. Maternal synchrony is more often associated with calming regulation and vagal efficiency, whereas paternal synchrony, frequently emerging in play-based or mildly arousing contexts, may be particularly relevant for behavioural regulation, exploratory competence, and adaptive engagement with novelty [166, 170, 190]. Disruptions in parental neurobiology, including altered oxytocin, cortisol, or autonomic functioning, are associated with reduced co-regulatory capacity and less optimal synchrony, particularly in the context of parental depression, anxiety, chronic stress, or histories of maltreatment [191–193].

Crucially, synchrony represents a modifiable pathway for intervention. High-quality caregiving relationships can buffer contextual risk at the level of physiological and behavioural synchrony, and growing evidence suggests that synchrony functions as a mechanism through which parenting-focused and relational interventions promote emotion regulation and reduce later psychopathology risk [194]. At the same time, methodological work cautions against simplistic interpretations of RSA as a direct proxy for vagal tone, underscoring the need to account for respiratory influences, situational context, and developmental stage when interpreting autonomic synchrony [195–197].

Taken together, parent-child synchrony and co-regulation provide a developmentally grounded and mechanistically plausible pathway linking parental mental health, early experience, and the calibration of regulatory systems central to brain health.

Section 6 therefore turns to the intervention implications, with an emphasis on timing-sensitive, context-embedded strategies for brain health promotion (see also Table 1).

## Implications for intervention and brain health promotion

Building directly on the life-course developmental frameworks outlined in Sections 4 and 5, this section considers how principles of developmental plasticity, biological embedding, and adaptive regulation can inform intervention and brain health promotion. Rather than emphasising risk reduction alone, a brain health perspective prioritises developmentally timed and context-sensitive strategies that support adaptive calibration of stress, affective, and social regulatory systems [127, 130, 198].

### Timing-sensitive intervention windows

Evidence across developmental neuroscience, psychiatry, and immune-psychiatry converges on prenatal life, childhood and adolescence as periods of heightened plasticity and opportunity. Interventions delivered across preconception, prenatal, and early postnatal windows, such as strengthening preconception and interconception health, perinatal mental health support, and early caregiving contexts, may yield durable benefits by shaping regulatory trajectories early across the life course, rather than relying on later remediation [28, 30, 199].

At the same time, evidence that aspects of plasticity can be partially re-engaged in adulthood cautions against deterministic interpretations of sensitive periods. Reopening plasticity without adequate environmental support may increase vulnerability rather than promote recovery, underscoring the need to align biological targets with psychosocial scaffolding [200].

### Family- and father-inclusive approaches

Findings reviewed in Section 5 indicate that parental mental health before, during, and after pregnancy influences offspring development through intergenerational and relational pathways. Clinical strategies focusing exclusively on mothers' risk overlooking modifiable paternal and family-level influences. Family-based, father-inclusive approaches that support parental regulation, caregiving quality, and co-regulation therefore represent high-leverage targets for promoting child brain health [127, 130]. Cross-cultural studies suggest that while the developmental relevance of peripartum stress and caregiving quality is broadly conserved, the form, meaning, and buffering resources of caregiving differ systematically across WEIRD and non-WEIRD contexts, with implications for intervention generalisability [66, 201].

### Targeting relational and co-regulatory mechanisms

Parent-child synchrony and co-regulation are modifiable mechanisms linking parental mental health to child outcomes. Interventions that enhance sensitive caregiving, dyadic regulation, and parent-child physiological synchrony may buffer the effects of ELA on stress and emotion-regulation systems. These approaches align with evidence that social buffering modulates neuroimmune and neuroplastic pathways implicated in psychopathology [200, 202].

### Precision, stratified, and clinically grounded intervention logic

In this review, *precision* refers to aligning intervention targets with developmentally and clinically meaningful phenomena, rather than biological specificity alone. Heterogeneity in developmental timing, adversity dimensions, sex, and biological embedding argues against one-size-fits-all approaches and motivates stratified interventions integrating clinical assessment with biological, behavioural, and contextual indicators [162, 199, 203]. From this perspective, precision psychiatry -and related advances in immune-psychiatry- are best understood as identifying modulators of plasticity, stress responsivity, and learning that vary across individuals and developmental stages, rather than as disorder-specific biomarker selection. Biological sophistication cannot compensate for poorly defined clinical phenomena: symptoms remain the primary entry point into care and must be interpreted as developmentally embedded and context-dependent [204].

Consistent with the RDoC-HiTOP framework outlined in Section 4.5, biological and digital markers are therefore most appropriately positioned as tools for stratification and hypothesis testing within dimensional, developmentally informed models. Quantifiable aspects of functioning, such as affective reactivity, autonomic regulation, and interpersonal dynamics, offer promising bridges between clinical observation and underlying neurobiological and immunological processes. When embedded within family- and context-sensitive frameworks, this precision logic can guide intervention timing and targets, such as combining prenatal stress reduction with postnatal co-regulation support, without presuming fixed categories or deterministic prediction.

Together, these implications motivate intervention research that is developmentally targeted and context-embedded. Section 7 therefore outlines priorities for future brain health science, focusing on study designs and measurement strategies needed to test timing-dependent mechanisms and support clinically interpretable translation.

## Future directions in brain health science

Brain health science is emerging as an integrative field linking neuroscience, psychiatry, public health, and social context to explain how developmental plasticity shapes lifelong functioning. A central challenge is to move from broad frameworks to testable, clinically interpretable models specifying when, how and for whom early experiences shift trajectories toward adaptive flexibility versus maladaptive rigidity [144, 145].

Future research should prioritise longitudinal, lifespan-oriented designs that explicitly test non-linear and timing-dependent effects of adversity and enrichment, while distinguishing empirically grounded adversity dimensions (Fig. 2D). Progress will depend on integrating multimodal biomarkers, including neuroimaging, autonomic indices, neuroimmune markers, and epigenetic profiles, to identify mechanistic signatures of calibration, sensitisation, and recovery [200].

A further priority is to strengthen translation by linking mechanistic measures to functional outcomes relevant to brain health broadly -such as cognition, emotion regulation, stress responsivity, social functioning, and aging-related vulnerability- rather than disorder endpoints alone. Big-data approaches combining multiomics, environmental exposure measures, and explainable machine learning may help detect complex interaction patterns, but require rigorous governance, transparent validation, and explicit attention to equity to ensure clinical relevance [199, 203, 205].

From a public-health and policy perspective, these priorities support expanding brain health promotion beyond prenatal care to include preconception and interconception health strategies, addressing stress regulation, mental health at the family and population level [122, 206–208].

Finally, multisystem resilience frameworks underscore that brain health is shaped not only by individual biology, but also by social, cultural, and institutional contexts. Equitable access to supportive environments, healthcare, and education remains a fundamental determinant of lifelong brain health. Together, these insights point toward a future in which harnessing developmental plasticity, neurosocial regulation, and contextual support can transform brain health trajectories across generations [127, 130].

**Author contributions** MC and BRHVdB conceptualized the study, MC, MGF, ETC, GASM, MCA, SDF, BRHvdB wrote the initial draft. MC and BRHVdB revised the final version. MC, MGF, ETC, GASM, MCA, SDF, BRHvdB approved the final version.

**Funding** MC and ETC are supported by University of Buenos Aires and CONICET. GASM is supported by the Dr. Bryan Kolb Chair in Neuroscience, University of Lethbridge. BRHVdB is supported by SBO project S003524N of the Research Foundation Flanders (Fonds voor Wetenschappelijk Onderzoek (FWO)-Vlaanderen) and by COST Action CA22114, funded by European Cooperation in Science and Technology.

**Competing interests** MGF holds US patents on maternal-foetal health monitoring and equity in several pregnancy health start-ups. The authors declare that there are no other competing interests associated with the manuscript.

**Ethics** During the first draft and the revisions, the authors used AI for literature search (Consensus, Endnote), language editing and improvement of clarity and coherence (ChatGPT). The tool was not used to generate scientific content or interpret evidence. All content, references, interpretations, and final wording were critically reviewed, edited, and approved by the authors.

## References

1. Reh RK, Dias BG, Nelson CA, Kaufer D, Werker JF, Kolb B, et al. Critical period regulation across multiple timescales. Proc Natl Acad Sci U A. 2020;117:23242–23251.
2. Faraji J, Metz GAS. The dynamic relationship of brain-derived neurotrophic factor and oxytocin: introducing the concept of neurosocial plasticity. Neuroprotection. 2025;3:63–78.
3. Avan A, Hachinski V. Brain health: key to health, productivity, and well-being. Alzheimer Dement. 2021;18:1396–1407.

4. Moguilner S, Baez S, Hernandez H, Migeot J, Legaz A, Gonzalez-Gomez R, et al. Brain clocks capture diversity and disparities in aging and dementia across geographically diverse populations. Nat Med. 2024;30:3646–3657.
5. Park C, Kim BR, Lim SM, Kim E-H, Jeong JH, Kim GH. Preserved brain youthfulness: longitudinal evidence of slower brain aging in superagers. GeroScience. 2025;47:5669–5679.
6. Chen Y, Demnitz N, Yamamoto S, Yaffe K, Lawlor B, Leroi I Defining brain health: a concept analysis. Int J Geriatr Psychiatry 2022; 37.
7. Seitz-Holland J, Haas SS, Penzel N, Reichenberg A, Pasternak O. BrainAGE, brain health, and mental disorders: a systematic review. Neurosci Biobehav Rev. 2024;159:105581.
8. Hachinski V. Toward a more inclusive definition of brain health. Neurology. 2023;101:580–581.
9. Vanderhaeghen P, Polleux F. Developmental mechanisms underlying the evolution of human cortical circuits. Nat Rev Neurosci. 2023;24:213–232.
10. Bethlehem RAI, Seidlitz J, White SR, Vogel JW, Anderson KM, Adamson C, et al. Brain charts for the human lifespan. Nature. 2022;604:525–533.
11. Hilberdink CE, van Zuiden M, Olff M, Roseboom TJ, de Rooij SR. The impact of adversities across the lifespan on psychological symptom profiles in late adulthood: a latent profile analysis. J Dev Orig Health Dis. 2023;14:508–522.
12. Walhovd KB, Krogsrud SK, Amlien IK, Sørensen Ø, Wang Y, Bråthen ACS, et al. Fetal influence on the human brain through the lifespan. eLife. 2024;12:RP86812.
13. McEwen BS. Protective and damaging effects of stress mediators. N Engl J Med. 1998;338:171–179.
14. Lupien SJ, McEwen BS, Gunnar MR, Heim C. Effects of stress throughout the lifespan on the brain, behaviour and cognition. Nat Rev Neurosci. 2009;10:434–445.
15. Slavich GM. Stressnology: the primitive (and problematic) study of life stress exposure and pressing need for better measurement. Brain Behav Immun. 2019;75:3–5.
16. Callaghan BL, Tottenham N. The stress acceleration hypothesis: effects of early-life adversity on emotion circuits and behavior. Curr Opin Behav Sci. 2016;7:76–81.
17. McEwen BS, Akil H. Revisiting the stress concept. J Neurosci. 2020;40:12–21.
18. Van den Bergh BRH, van den Heuvel MI, Lahti M, Braeken M, de Rooij SR, Entringer S, et al. Prenatal developmental origins of behavior and mental health: The influence of maternal stress in pregnancy. Neurosci Biobehav Rev. 2020;117:26–64.
19. Fisher SD, Cobo J, Figueiredo B, Fletcher R, Garfield CF, Hanley J, et al. Expanding the international conversation with fathers' mental health: toward an era of inclusion in perinatal research and practice. Arch Womens Ment Health. 2021;24:841–848.
20. Van den Bergh BRH, Antonelli MC, Stein DJ. Current perspectives on perinatal mental health and neurobehavioral development: focus on regulation, coregulation and self-regulation. Curr Opin Psychiatry. 2024;37:237–250.
21. Cánepa ET, Berardino BG. Epigenetic mechanisms linking early-life adversities and mental health. Biochem J. 2024;481:615–642.
22. Ambeskovic M, Babenko O, Ilnytskyy Y, Kovalchuk I, Kolb B, Metz GAS. Ancestral stress alters lifetime mental health trajectories and cortical neuromorphology via epigenetic regulation. Sci Rep. 2019;9:6389.
23. Hubel DH, Wiesel TN. The period of susceptibility to the physiological effects of unilateral eye closure in kittens. J Physiol. 1970;206:419–436.
24. Greenough WT, Black JE, Wallace CS. Experience and brain development. Child Dev. 1987;58:539–559.
25. Hensch TK Critical period mechanisms in developing visual cortex. In: Current Topics in Developmental Biology. 2005, pp. 215–237.
26. Nelson CA, Gabard-Durnam LJ. Early adversity and critical periods: neurodevelopmental consequences of violating the expectable environment. Trends Neurosci. 2020;43:133–143.
27. Bourgeois J. Synaptogenesis, heterochrony and epigenesis in the mammalian neocortex. Acta Paediatr. 1997;86:27–33.
28. Knudsen EI. Sensitive periods in the development of the brain and behavior. J Cogn Neurosci. 2004;16:1412–1425.
29. Teicher MH, Samson JA. Annual research review: enduring neurobiological effects of childhood abuse and neglect. J Child Psychol Psychiatry. 2016;57:241–66.
30. Gabard-Durnam LJ, McLaughlin KA. Sensitive periods in human development. Curr Opin Behav Sci. 2019;28:69–75.
31. Larsen B, Luna B. Adolescence as a neurobiological critical period for the development of higher-order cognition. Neurosci Biobehav Rev. 2018;94:179–195.
32. Perica MI, Calabro FJ, Larsen B, Foran W, Yushmanov VE, Hetherington H, et al. Development of frontal GABA and glutamate supports excitation/inhibition balance from adolescence into adulthood. Prog Neurobiol. 2022;219:102370.
33. Hensch TK, Fagiolini M. Excitatory-inhibitory balance and critical period plasticity in developing visual cortex. Prog Brain Res. 2005;147:115–124.
34. Nabel EM, Morishita H. Regulating critical period plasticity: Insight from the visual system to fear circuitry for therapeutic interventions. Front Psychiatry. 2013;4:146.
35. Carulli D, Verhaagen J. An extracellular perspective on cns maturation: Perineuronal nets and the control of plasticity. Int J Mol Sci. 2021;22:2434.
36. Guo W, Wang X, Zhou Z, Li Y, Hou Y, Wang K, et al. Advances in fear memory erasure and its neural mechanisms. Front Neurol 2024. 2025;15:1481450.

37. Putignano E, Lonetti G, Cancedda L, Ratto G, Costa M, Maffei L, et al. Developmental downregulation of histone posttranslational modifications regulates visual cortical plasticity. Neuron. 2007;53:747–59.
38. Gu Y, Tran T, Murase S, Borrell A, Kirkwood A, Quinlan EM. Neuregulin-dependent regulation of fast-spiking interneuron excitability controls the timing of the critical period. J Neurosci. 2016;36:10285–10295.
39. Sun Y, Ikrar T, Davis MF, Gong N, Zheng X, Luo ZD, et al. Neuregulin-1/ErbB4 signaling regulates visual cortical plasticity. Neuron. 2016;92:160–173.
40. Duncan BW, Murphy KE, Maness PF. Molecular mechanisms of L1 and NCAM adhesion molecules in synaptic pruning, plasticity, and stabilization. Front Cell Dev Biol. 2021;9:625340.
41. Starkey J, Horstick EJ, Ackerman SD. Glial regulation of critical period plasticity. Front Cell Neurosci. 2023;17:1247335. https://doi.org/10.3389/FNCEL.2023.1247335
42. Vivi E, Di Benedetto B. Brain stars take the lead during critical periods of early postnatal brain development: relevance of astrocytes in health and mental disorders. Mol Psychiatry. 2024;29:2821–2833.
43. Horn ME, Nicoll RA. Somatostatin and parvalbumin inhibitory synapses onto hippocampal pyramidal neurons are regulated by distinct mechanisms. Proc Natl Acad Sci USA. 2018;115:589–594.
44. Hunter I, Coulson B, Pettini T, Davies JJ, Parkin J, Landgraf M, et al. Balance of activity during a critical period tunes a developing network. eLife. 2024;12:RP91599. https://doi.org/10.7554/ELIFE.91599
45. Poplawski J, Montina T, Metz GAS. Early life stress shifts critical periods and causes precocious visual cortex development. PLoS ONE. 2024;19:e0316384.
46. Berardino BG, Chertoff M, Gianatiempo O, Alberca CD, Priegue R, Fiszbein A, et al. Exposure to enriched environment rescues anxiety-like behavior and miRNA deregulated expression induced by perinatal malnutrition while altering oligodendrocyte morphology. Neuroscience. 2019;408:115–134.
47. Margolis ET, Gabard-Durnam LJ. Prenatal influences on postnatal neuroplasticity: Integrating DOHaD and sensitive/critical period frameworks to understand biological embedding in early development. Infancy. 2025;30:e12588.
48. Alberca CD, Papale LA, Madrid A, Gianatiempo O, Cánepa ET, Alisch RS, et al. Perinatal protein malnutrition results in genome-wide disruptions of 5-hydroxymethylcytosine at regions that can be restored to control levels by an enriched environment. Epigenetics. 2021;16:1085–1101.
49. Felitti VJ, Anda RF, Nordenberg D, Williamson DF, Spitz AM, Edwards V, et al. Relationship of childhood abuse and household dysfunction to many of the leading causes of death in adults: the adverse childhood experiences (ACE) Study. Am J Prev Med. 1998;14:245–258.
50. Letcher P, Greenwood CJ, McAnally H, Belsky J, Macdonald JA, Spry EA, et al. Parental history of positive development and child behavior in next generation offspring: a two-cohort prospective intergenerational study. Child Dev. 2023;94:60–73.
51. McLaughlin KA, Sheridan MA. Beyond cumulative risk: a dimensional approach to childhood adversity. Curr Dir Psychol Sci. 2016;25:239–245.
52. De Asis-Cruz J, Andescavage N, Limperopoulos C. Adverse prenatal exposures and fetal brain development: insights from advanced fetal magnetic resonance imaging. Biol Psychiatry Cogn Neurosci Neuroimaging. 2022;7:480–490.
53. De Asis-Cruz J, Krishnamurthy D, Zhao L, Kapse K, Vezina G, Andescavage N, et al. Association of prenatal maternal anxiety with fetal regional brain connectivity. JAMA Netw Open. 2020;3:e2022349.
54. Hay RE, Reynolds JE, Grohs MN, Paniukov D, Giesbrecht GF, Letourneau N, et al. Amygdala-prefrontal structural connectivity mediates the relationship between prenatal depression and behavior in preschool boys. J Neurosci. 2020;40:6969–6977.
55. Marečková K, Klasnja A, Andrýsková L, Brázdil M, Paus T. Developmental origins of depression-related white matter properties: findings from a prenatal birth cohort. Hum Brain Mapp. 2019;40:1155–1163.
56. Van den Bergh BRH, Sleurs C, Geusens B, Emsell L, Sunaert S, Billiet T. White matter microstructure and cognitive abilities in 28-year-old offspring prenatally exposed to maternal anxiety: a prospective exploratory multimodal brain imaging study. Brain Cogn. 2025;188:106319.
57. Vinogradov S, Chafee MV, Lee E, Morishita H. Psychosis spectrum illnesses as disorders of prefrontal critical period plasticity. Neuropsychopharmacology. 2023;48:168–185.
58. Zinn ME, Huntley ED, Keating DP. Resilience in adolescence: prospective self moderates the association of early life adversity with externalizing problems. J Adolesc. 2020;81:61–72.
59. Zhang D, Fang J, Zhang L, Yuan J, Wan Y, Su P, et al. Pubertal recalibration of cortisol reactivity following early life parent-child separation. J Affect Disord. 2021;278:320–326.
60. Teicher MH, Gordon JB, Nemeroff CB. Recognizing the importance of childhood maltreatment as a critical factor in psychiatric diagnoses, treatment, research, prevention, and education. Mol Psychiatry. 2022;27:1331–1338.
61. Farah MJ. The neuroscience of socioeconomic status: correlates, causes, and consequences. Neuron. 2017;96:56–71.
62. Johnson SB, Riis JL, Noble KG. State of the art review: poverty and the developing brain. Pediatrics. 2016;137:e20153075.
63. Berardino BG, Cardillo MB, Priegue R, Perez HC, Nemirovsky SI, Salvochea M, et al. A novel multidimensional, perinatal social and material deprivation paradigm induces aggressive behavior and cortical gene expression reprogramming in young adult mice. Prog Neuropsychopharmacol Biol Psychiatry. 2025;140:111432.
64. Henrich J, Heine SJ, Norenzayan A. The weirdest people in the world? Behav Brain Sci. 2010;33:61–83.

65. MacGinty R, Hoffman N, Koen N, Zar HJ, Stein DJ, Pellowski J. Longitudinally investigating patterns of maternal psychological distress in a South African birth cohort. BMC Public Health. 2025;25:3409.
66. Burger M, Hoosain M, Einspieler C, Unger M, Niehaus D. Maternal perinatal mental health and infant and toddler neurodevelopment - evidence from low and middle-income countries. a systematic review. J Affect Disord. 2020;268:158–172.
67. Buffa G, Dahan S, Sinclair I, St-Pierre M, Roofigari N, Mutran D, et al. Prenatal stress and child development: a scoping review of research in low- and middle-income countries. PloS One. 2018;13:e0207235.
68. van der Heijden F, Ali N, van den Heuvel MI, De Baene W, Sitskoorn M. The interplay between poverty and the human brain connectome across the lifespan: a systematic review. Neurosci Biobehav Rev. 2026;181:106532.
69. Noble KG, Houston SM, Brito NH, Bartsch H, Kan E, Kuperman JM, et al. Family income, parental education and brain structure in children and adolescents. Nat Neurosci. 2015;18:773–778.
70. Bush NR, Edgar RD, Park M, MacIsaac JL, McEwen LM, Adler NE, et al. The biological embedding of early-life socioeconomic status and family adversity in children's genome-wide DNA methylation. Epigenomics. 2018;10:1445–1461.
71. Chae DH, Snipes SA, Chung KW, Martz CD, LaVeist TA. Vulnerability and resilience: use and misuse of these terms in the public health discourse. Am J Public Health 1971. 2021;111:1736–1740.
72. Bock J, Rether K, Gröger N, Xie L, Braun K. Perinatal programming of emotional brain circuits: an integrative view from systems to molecules. Front Neurosci. 2014;8:1–16.
73. Mcewen BS, Nasca C, Gray JD. Stress effects on neuronal structure: hippocampus, amygdala, and prefrontal cortex. Neuropsychopharmacol Off Publ Am Coll Neuropsychopharmacol. 2016;41:3–23.
74. Chan SY, Ngoh ZM, Ong ZY, Teh AL, Kee MZL, Zhou JH, et al. The influence of early-life adversity on the coupling of structural and functional brain connectivity across childhood. Nat Ment Health. 2024;2:52–62.
75. Grillo Balboa J, Colapietro AA, Cantarelli VI, Ponzio MF, Ceol Retamal MN, Pallarés ME, et al. Sex-specific outcomes in a rat model of early-life stress due to adverse caregiving. Neurotox Res. 2025;43:10.
76. Bock J, Gruss M, Becker S, Braun K. Experience-induced changes of dendritic spine densities in the prefrontal and sensory cortex: correlation with developmental time windows. Cereb Cortex N Y N 1991. 2005;15:802–808.
77. Braun K, Bock J. Jör. The experience-dependent maturation of prefronto-limbic circuits and the origin of developmental psychopathology: implications for the pathogenesis and therapy of behavioural disorders. Dev Med Child Neurol. 2011;53:14–18.
78. Antonow-Schlorke I, Schwab M, Li C, Nathanielsz PW. Glucocorticoid exposure at the dose used clinically alters cytoskeletal proteins and presynaptic terminals in the fetal baboon brain. J Physiol. 2003;547:117–23.
79. Baram TZ, Birnie MT. Enduring memory consequences of early-life stress / adversity: Structural, synaptic, molecular and epigenetic mechanisms. Neurobiol Stress. 2024;33:100669.
80. Birnie MT, Baram TZ. The evolving neurobiology of early-life stress. Neuron. 2025;113:1474–1490.
81. McGowan PO, Sasaki A, D'Alessio AC, Dymov S, Labonté B, Szyf M, et al. Epigenetic regulation of the glucocorticoid receptor in human brain associates with childhood abuse. Nat Neurosci. 2009;12:342–348.
82. Weaver ICG, Cervoni N, Champagne FA, D'Alessio AC, Sharma S, Seckl JR, et al. Epigenetic programming by maternal behavior. Nat Neurosci. 2004;7:847–854.
83. Vaiserman AM, Koliada AK. Early-life adversity and long-term neurobehavioral outcomes: epigenome as a bridge? Hum Genomics. 2017;11:34.
84. Dion A, Muñoz PT, Franklin TB. Epigenetic mechanisms impacted by chronic stress across the rodent lifespan. Neurobiol Stress. 2022;17:100434.
85. Gerasymchuk M, Cherkasova V, Kovalchuk O, Kovalchuk I. The role of microRNAs in organismal and skin aging. Int J Mol Sci. 2020;21:5281.
86. Xu P, Ma Y, Wu H, Wang Y-L. Placenta-Derived MicroRNAs in the pathophysiology of human pregnancy. Front Cell Dev Biol. 2021;9:646326.
87. Marsland AL, Gianaros PJ, Kuan DC-H, Sheu LK, Krajina K, Manuck SB. Brain morphology links systemic inflammation to cognitive function in midlife adults. Brain Behav Immun. 2015;48:195–204.
88. Frodl T, Amico F. Is there an association between peripheral immune markers and structural/functional neuroimaging findings? Prog Neuropsychopharmacol Biol Psychiatry. 2014;48:295–303.
89. Hyon R, Youm Y, Kim J, Chey J, Kwak S, Parkinson C. Similarity in functional brain connectivity at rest predicts interpersonal closeness in the social network of an entire village. Proc Natl Acad Sci USA. 2020;117:33149–33160.
90. Meisner OC, Nair A, Chang SWC. Amygdala connectivity and implications for social cognition and disorders. Handb Clin Neurol. 2022;187:381–403.
91. Maynard KR, Hobbs JW, Phan BN, Gupta A, Rajpurohit S, Williams C, et al. BDNF-TrkB signaling in oxytocin neurons contributes to maternal behavior. eLife. 2018;7:e33676.
92. Mitre M, Saadipour K, Williams K, Khatri L, Froemke RC, Chao MV. Transactivation of TrkB receptors by oxytocin and its G protein-coupled receptor. Front Mol Neurosci. 2022;15:891537.
93. Faraji J, Lotfi H, Moharrerie A, Jafari SY, Soltanpour N, Tamannaiee R, et al. Regional differences in BDNF expression and behavior as a function of sex and enrichment type: oxytocin matters. Cereb Cortex. 2022;32:2985–2999.

94. Mansouri M, Pouretemad H, Roghani M, Wegener G, Ardalan M. Autistic-like behaviours and associated brain structural plasticity are modulated by oxytocin in maternally separated rats. Behav Brain Res. 2020;393:112756.
95. Dunn EC, Soare TW, Zhu Y, Simpkin AJ, Suderman MJ, Klengel T, et al. Sensitive periods for the effect of childhood adversity on DNA methylation: results from a prospective, longitudinal study. Biol Psychiatry. 2019;85:838–849.
96. Calabrese EJ, Baldwin LA. Hormesis: U-shaped dose responses. Trends Pharmacol Sci. 2001;22:285–291.
97. Agathokleous E, Calabrese EJ. Editorial overview: hormesis and dose-response. Curr Opin Toxicol. 2022;30:100343.
98. Mattson MP. Challenging oneself intermittently to improve health. Dose-Response. 2014;12:600–618.
99. Calabrese EJ, Mattson MP. The catabolic - anabolic cycling hormesis model of health and resilience. Ageing Res Rev. 2024;102:102588.
100. Stranahan AM, Mattson MP. Recruiting adaptive cellular stress responses for successful brain ageing. Nat Rev Neurosci. 2012;13:209–216.
101. van Praag H, Fleshner M, Schwartz MW, Mattson MP. Exercise, energy intake, glucose homeostasis, and the brain. J Neurosci. 2014;34:15139–15149.
102. Oshri A, Howard CJ, Zhang L, Reck A, Cui Z, Liu S, et al. Strengthening through adversity: the hormesis model in developmental psychopathology. Dev Psychopathol. 2024;36:2390–2406.
103. Rutter M. Resilience as a dynamic concept. Dev Psychopathol. 2012;24:335–344.
104. Hanson MA, Gluckman PD. Early developmental conditioning of later health and disease: physiology or pathophysiology? Physiol Rev. 2014;94:1027–76.
105. Van den Bergh BRH. Developmental programming of early brain and behaviour development and mental health: a conceptual framework. Dev Med Child Neurol. 2011;53:19–23.
106. Monk C, Fernández CR. Neuroscience advances and the developmental origins of health and disease research. JAMA Netw Open. 2022;5:e229251.
107. Sroufe LA, Rutter M. The domain of developmental psychopathology. Child Dev. 1984;55:17–29.
108. Cicchetti D, Rogosch FA. Psychopathology as risk for adolescent substance use disorders: a developmental psychopathology perspective. J Clin Child Psychol. 1999;28:355–365.
109. Cicchetti D. Annual research review: resilient functioning in maltreated children-past, present, and future perspectives. J Child Psychol Psychiatry. 2013;54:402–22.
110. Ioannidis K, Askelund AD, Kievit RA, van Harmelen AL. The complex neurobiology of resilient functioning after childhood maltreatment. BMC Med. 2020;18:32.
111. Van den Bergh BRH, Mulder EJH, Mennes M, Glover V. Antenatal maternal anxiety and stress and the neurobehavioural development of the fetus and child: links and possible mechanisms. a review. Neurosci Biobehav Rev. 2005;29:237–258.
112. Van den Bergh BRH, Mulder EJH, Visser GHA, Poelmann-Weesjes G, Bekedam DJ, Prechtl HFR. The effect of (induced) maternal emotions on fetal behaviour: a controlled study. Early Hum Dev. 1989;19:9–19.
113. Ferreira AJ. Emotional factors in prenatal environment a review. J Nerv Ment Dis. 1965;141:108–118.
114. Carlson DB, Labarba RC. Maternal emotionality during pregnancy and reproductive outcome: a review of the literature. Int J Behav Dev. 1979;2:343–376.
115. Stott DH. Physical and mental handicaps following a disturbed pregnancy. The Lancet. 1957;269:1006–1012.
116. Sontag LW. The significance of fetal environmental differences. Am J Obstet Gynecol. 1941;42:996–1003.
117. Joffe JM. Prenatal determinants of behaviour. Oxford: Pergamon Press; 1969. Oxford, 1969
118. Osofsky JD (ed.). Handbook of infant development. New York; Wiley: 1979.
119. Del Giudice M, Ellis BJ, Shirtcliff EA. The adaptive calibration model of stress responsivity. Neurosci Biobehav Rev. 2011;35:1562–1592.
120. Del Giudice M. Early stress and human behavioral development: emerging evolutionary perspectives. J Dev Orig Health Dis. 2014;5:270–280.
121. Lu MC, Halfon N. Racial and ethnic disparities in birth outcomes: a life-course perspective. Matern Child Health J. 2003;7:13–30.
122. Halfon N, Forrest CB The emerging theoretical framework of life course health development. In: Halfon N, Forrest CB, Lerner RM, Faustman EM (eds). Handbook of Life Course Health Development. Cham; Springer International Publishing: 2018, pp. 19–43.
123. Rutter M. Psychosocial resilience and protective mechanisms. Am J Orthopsychiatry. 1987;57:316–331.
124. Meichenbaum D A self-instructional approach to stress management. Stress Anxiety Mod Life 1976.
125. Seery MD, Holman EA, Silver RC. Whatever does not kill us. J Pers Soc Psychol. 2010;99:1025–1041.
126. Gunnar MR, Frenn K, Wewerka SS, Van Ryzin MJ. Moderate versus severe early life stress. Psychoneuroendocrinology. 2009;34:62–75.
127. Masten AS. Resilience in development and psychopathology. Annu Rev Clin Psychol. 2021;17:521–549.
128. Bethell CD, Solloway MR, Guinosso S, Hassink S, Srivastav A, Ford D, et al. Prioritizing possibilities for child and family health: an agenda to address adverse childhood experiences and foster the social and emotional roots of well-being in pediatrics. Acad Pediatr. 2017;17:S36–S50.
129. Bethell C, Jones J, Gombojav N, Linkenbach J, Sege R. Positive childhood experiences and adult mental and relational health in a statewide sample. JAMA Pediatr. 2019;173:e193007.
130. Kalisch R. The resilience framework. Nat Hum Behav. 2017;1:784–790.

131. Bonanno GA, Chen S, Galatzer-Levy IR. Resilience to potential trauma and adversity through regulatory flexibility. Nat Rev Psychol. 2023;2:663–675.
132. Bonanno GA, Chen S, Bagrodia R, Galatzer-Levy IR. Resilience and disaster: flexible adaptation in the face of uncertain threat. Annu Rev Psychol. 2024;75:573–599.
133. Kalisch R, Russo SJ, Müller MB. Neurobiology and systems biology of stress resilience. Physiol Rev. 2024;104:1205–1263.
134. Antonovsky A Unraveling the mystery of health: How people manage stress and stay well. Jossey-Bass. Jossey-Bass, 1987.
135. Mittelmark MB Salutogenesis from its origins to the present. In: Mittelmark MB, Bauer GF, Vaandrager L, Pelikan JM, Sagy S, Eriksson M et al. (eds). The Handbook of Salutogenesis. Cham (CH); 2022, pp. 3–4.
136. Hartman S, Belsky J, Pluess M. Prenatal programming of environmental sensitivity. Transl Psychiatry. 2023;13:161.
137. Belsky J, Bakermans-kranenburg MJ, van IJzendoorn MH. MHV. for better and for worse: differential susceptibility to environmental influences. Curr Dir Psychol Sci. 2007;16:300–304.
138. Friston K. The free-energy principle: a unified brain theory? Nat Rev Neurosci. 2010;11:127–138.
139. Clark A. Whatever next? Predictive brains. Behav Brain Sci. 2013;36:181–204.
140. Seth AK. Interoceptive inference, emotion, and the embodied self. Trends Cogn Sci. 2013;17:565–573.
141. Barrett LF, Simmons WK. Interoceptive predictions in the brain. Nat Rev Neurosci. 2015;16:419–429.
142. Desmedt O, Van den Bergh O. Beyond interoceptive accuracy: new directions in interoception research. Biol Psychol. 2024;189:108800.
143. Sauer KS, Pohl A, Van den Bergh O, Witthöft M. Support for predictive processing assumptions on health anxiety: imprecise expectations on the origin of somatic symptoms increase health anxiety. Biopsychosoc Sci Med. 2025;87:672–681.
144. Bowman H. Is predictive coding falsifiable? Neurosci Biobehav Rev. 2023;154:105404.
145. Hodson R, Mehta M, Smith R. The empirical status of predictive coding and active inference. Neurosci Biobehav Rev. 2024;157:105473.
146. Mangalam M. The myth of the bayesian brain. Eur J Appl Physiol. 2025;125:2643–2677.
147. Hyman SE. The diagnosis of mental disorders: the problem of reification. Annu Rev Clin Psychol. 2010;6:155–179.
148. Kendler KS. The nature of psychiatric disorders. World Psychiatry. 2016;15:5–12.
149. Zautra N. Psychiatry's new validity crisis: the problem of disparate validation. Philos Sci. 2025;92:646–665.
150. Morris SE, Sanislow CA, Pacheco J, Vaidyanathan U, Gordon JA, Cuthbert BN. Revisiting the seven pillars of RDoC. BMC Med. 2022;20:220.
151. Cuthbert BN. Research domain criteria (RDoC): progress and potential. Curr Dir Psychol Sci. 2022;31:107–114.
152. McLaughlin KA, Gabard-Durnam L. Experience-driven plasticity and the emergence of psychopathology: a mechanistic framework integrating development and the environment into the research domain criteria (RDoC) model. J Psychopathol Clin Sci. 2022;131:575–587.
153. Wade M, Carroll D, Fox NA, Zeanah CH, Nelson CA. Associations between early psychosocial deprivation, cognitive and psychiatric morbidity, and risk-taking behavior in adolescence. J Clin Child Adolesc Psychol. 2022;51:850–863.
154. Ellis BJ, Sheridan MA, Belsky J, McLaughlin KA. Why and how does early adversity influence development? Toward an integrated model of dimensions of environmental experience. Dev Psychopathol. 2022;34:447–471.
155. Barrantes-Vidal N, Torrecilla P, Lavín V, Mas-Bermejo P, Papiol S, Bakermans-Kranenburg MJ, et al. Time to incorporate positive experiences and outcomes in psychosis research: genetic differential susceptibility to childhood experiences works for 'better and for worse. Transl Psychiatry. 2025;15:439.
156. Kotov R, Cicero DC, Conway CC, DeYoung CG, Dombrovski A, Eaton NR, et al. The hierarchical taxonomy of psychopathology (HiTOP) in psychiatric practice and research. Psychol Med. 2022;52:1666–1678.
157. Hagerty SL. Toward precision characterization and treatment of psychopathology: a path forward and integrative framework of the hierarchical taxonomy of psychopathology and the research domain criteria. Perspect Psychol Sci. 2023;18:91–109.
158. Davis CN, Khan Y, Toikumo S, Jinwala Z, Boomsma DI, Levey DF, et al. Integrating HiTOP and RDoC frameworks Part I: genetic architecture of externalizing and internalizing psychopathology. Psychol Med. 2025;55:e138.
159. Davis CN, Khan Y, Toikumo S, Jinwala Z, Boomsma DI, Levey DF, et al. Integrating HiTOP and RDoC frameworks part II: shared and distinct biological mechanisms of externalizing and internalizing psychopathology. Psychol Med. 2025;55:e137.
160. Del Casale A. The era of precision psychiatry: toward a new paradigm for multidimensional data integration in mental health. Assess Summ Verify Exact J. 2025;6:146.
161. Holt DJ, Choi KW, DeTore NR, Balogun O. Transdiagnostic prevention in youth mental health, part I: rationale, shared risk factors. Neuropsychopharmacology. 2026;51:293–309.
162. Kas M. The precision psychiatry roadmap: towards a biology-informed framework for mental disorders. Eur Psychiatry. 2025;68:S8–S9.
163. Halfon N, Forrest CB, Lerner RM, Faustman EM (eds). Handbook of Life Course Health Development. Cham; Springer International Publishing: 2018. https://doi.org/10.1007/978-3-319-47143-3.
164. Liu Q, Zhu S, Zhou X, Liu F, Becker B, Kendrick KM, et al. Mothers and fathers show different neural synchrony with their children during shared experiences. NeuroImage. 2024;288:120529.

165. Puglisi N, Favez N, Rattaz V, Epiney M, Razurel C, Tissot H. Interactive synchrony and infants' vagal tone as an index of emotion regulation: associations within each mother- and father-infant dyad and across dyads. Front Psychol. 2023;14:1299041.
166. Lunkenheimer E, Brown KM, Fuchs A. Differences in mother-child and father-child RSA synchrony: moderation by child self-regulation and dyadic affect. Dev Psychobiol. 2021;63:1210–1224.
167. Daneshnia N, Chechko N, Nehls S. Do parental hormone levels synchronize during the prenatal and postpartum periods? A systematic review. Clin Child Fam Psychol Rev. 2024;27:658–676.
168. McDonnell CG, Valentino K. Intergenerational effects of childhood trauma: evaluating pathways among maternal ACEs, perinatal depressive symptoms, and infant outcomes. Child Maltreat. 2016;21:317–326.
169. Uy JP, Tan AP, Broeckman BBFP, Gluckman PD, Chong YS, Chen H, et al. Effects of maternal childhood trauma on child emotional health: maternal mental health and frontoamygdala pathways. J Child Psychol Psychiatry. 2023;64:426–436.
170. Sobral M, Pacheco F, Perry B, Antunes J, Martins S, Guiomar R, et al. Neurobiological correlates of fatherhood during the postpartum period: a scoping review. Front Psychol. 2022;13:745767.
171. Silver J, Thorpe D, Olino TM, Klein DN. intergenerational effects of parenting on children's internalizing and externalizing trajectories: a latent growth model analysis. Child Psychiatry Hum Dev. 2025;56:312–327.
172. Klimek M, Karlsson H, Karlsson L, Korja R, Nolvi S, Häikiö T, et al. Paternal adverse childhood experiences and offspring's attentional disengagement from faces at 8 months-results from the FinnBrain Birth Cohort Study. PloS One. 2025;20:e0326437.
173. Le Bas G, Aarsman SR, Rogers A, Macdonald JA, Misuraca G, Khor S, et al. Paternal perinatal depression, anxiety, and stress and child development: a systematic review and meta-analysis. JAMA Pediatr. 2025;179:903–917.
174. Atzil S, Hendler T, Zagoory-Sharon O, Winetraub Y, Feldman R. Synchrony and specificity in the maternal and the paternal brain: relations to oxytocin and vasopressin. J Am Acad Child Adolesc Psychiatry. 2012;51:798–811.
175. Snippe-Strauss M, Tenenhaus-Zamir A, Benhos A, Richter-Levin G. Active stress resilience. Curr Opin Behav Sci. 2024;58:101403.
176. Phua DY, Chew CSM, Tan YL, Ng BJK, Lee FKL, Tham MMY. Differential effects of prenatal psychological distress and positive mental health on offspring socioemotional development from infancy to adolescence: a meta-analysis. Front Pediatr. 2023;11:1221232.
177. McCreary JK, Metz GAS. Environmental enrichment as an intervention for adverse health outcomes of prenatal stress. Environ Epigenetics. 2016;2:dvw013.
178. Eriksson M The sense of coherence: the concept and its relationship to health. Springer International Publishing, 2022, pp 61–68.
179. Lähdepuro A, Räikkönen K, Pham H, Thompson-Felix T, Eid RS, O'Connor TG, et al. Maternal social support during and after pregnancy and child cognitive ability: examining timing effects in two cohorts. Psychol Med. 2024;54:1661–1670.
180. Lähdepuro A, Lahti-Pulkkinen M, Pyhälä R, Tuovinen S, Lahti J, Heinonen K, et al. Positive maternal mental health during pregnancy and mental and behavioral disorders in children: a prospective pregnancy cohort study. J Child Psychol Psychiatry. 2023;64:807–816.
181. Suveg C, Braunstein West K, Davis M, Caughy M, Smith EP, Oshri A. Symptoms and synchrony: Mother and child internalizing problems moderate respiratory sinus arrhythmia concordance in mother-preadolescent dyads. Dev Psychol. 2019;55:366–376.
182. Mousa O, Alshakhsi S, Yankouskaya A, Panourgia C, Ali R Good technology requires a good environment: the role of parenting practices in adolescent internet addiction. 2024;143-150.
183. Tandon T. Mental health markers and protective factors in students with symptoms of physical pain across WEIRD and Non-weird samples – a network analysis. BMC Psychiatry. 2024;24:318.
184. Stein B, Hoeft F, Richter CG. Stress, resilience, and emotional well-being in children and adolescents with specific learning disabilities. Curr Opin Behav Sci. 2024;58:101410.
185. Gunnar M, Quevedo K. The neurobiology of stress and development. Annu Rev Psychol. 2007;58:145–173.
186. Feldman R. The neurobiology of human attachments. Trends Cogn Sci. 2017;21:80–99.
187. Abney DH, daSilva EB, Bertenthal BI. Associations between infant-mother physiological synchrony and 4- and 6-month-old infants' emotion regulation. Dev Psychobiol. 2021;63:e22161.
188. Leclère C, Viaux S, Avril M, Achard C, Chetouani M, Missonnier S, et al. Why synchrony matters during mother-child interactions: a systematic review. PloS One. 2014;9:e113571.
189. Stern JA, Kelsey CM, Yancey H, Grossmann T. Love on the developing brain: maternal sensitivity and infants' neural responses to emotion in the dorsolateral prefrontal cortex. Dev Sci. 2024;27:e13497.
190. Kim P, Leckman JF, Mayes LC, Feldman R, Wang X, Swain JE. The plasticity of human maternal brain: Longitudinal changes in brain anatomy during the early postpartum period. Behav Neurosci. 2010;124:695–700.
191. Lan Q, Zhang C, Lunkenheimer E, Chang S, Li Z, Wang L. Maternal postnatal depressive symptoms and children's internalizing problems: the moderating role of mother-infant RSA synchrony. Dev Psychopathol. 2024;36:1776–1788.
192. Brown SM, Lunkenheimer E, LeBourgeois M, Heilman K, et al. Maltreatment severity and sleep variability in toddlers predict RSA synchrony: findings from caregiver-child dyads. Dev Psychopathol. 2021;33:1747–1758.
193. Fuchs A, Lunkenheimer E, Brown K. Parental history of childhood maltreatment and child average RSA shape parent-child RSA synchrony. Dev Psychobiol. 2021;63:e22171.

194. Gray SAO, Miller JG, Glackin EB, Hatch V, Drury SS. Parent–child relationship quality buffers the association between mothers' adverse childhood experiences and physiological synchrony. J Child Psychol Psychiatry. 2025;66:956–966.
195. Grossman P. Respiratory sinus arrhythmia (RSA), vagal tone and biobehavioral integration: beyond parasympathetic function. Biol Psychol. 2024;186:108739.
196. Menuet C, Ben-Tal A, Linossier A, Allen AM, Machado BH, Moraes DJA, et al. Redefining respiratory sinus arrhythmia as respiratory heart rate variability: an international expert recommendation for terminological clarity. Nat Rev Cardiol. 2025;22:978–984.
197. Grossman P, Taylor EW. Toward understanding respiratory sinus arrhythmia: relations to cardiac vagal tone, evolution and biobehavioral functions. Biol Psychol. 2007;74:263–285.
198. Masten AS, Tyrell FA, Cicchetti D. Resilience in development: pathways to multisystem integration. Dev Psychopathol. 2023;35:2103–2112.
199. Noble KG, Magnuson K, Gennetian LA. Biological embedding of socioeconomic status through early childhood. Proc Natl Acad Sci. 2021;118:e2106730118.
200. Juster R-P, McEwen BS, Lupien SJ. Advancing the allostatic load model: a contemporary review and future directions. Psychoneuroendocrinology. 2023;152:106052.
201. Huidina J, Yamsani S. Folk wisdom and traditional knowledge: a study of mother and child healthcare practices of maram Nagas in India. J Asian Afr Stud. 2024;61:301–315.
202. Miller AH, Raison CL. The role of inflammation in depression: from evolutionary imperative to modern treatment target. Nat Rev Immunol. 2016;16:22–34.
203. Duncan GJ, Magnuson K, Votruba-Drzal E. Moving beyond correlations in assessing the consequences of poverty. Annu Rev Psychol. 2017;68:413–434.
204. Amad A, D'Hondt F, Daoudi M, Fovet T. No precision psychiatry without clinical precision. Mol Psychiatry. 2025;31:1171–1172.
205. Bracher-Smith M, Escott-Price V. Decoding the genomic symphony: unravelling brain disorders through data integration and machine learning. Mol Psychiatry. 2025;30:5914–5925.
206. Merz EC, Myers B, Hansen M, Simon KR, Strack J, Noble KG. Socioeconomic disparities in hypothalamic-pituitary-adrenal axis regulation and prefrontal cortical structure. Biol Psychiatry Glob Open Sci. 2024;4:83–96.
207. Shonkoff JP, Boyce WT, Levitt P, Martinez FD, McEwen B. Leveraging the biology of adversity and resilience to transform pediatric practice. Pediatrics. 2021;147:e20193845.
208. Framework to implement a life course approach in practice. Geneva: World Health Organization; 2025. Licence: CC BY-NC-SA 3.0 IGO.
209. Davenport MH, McCurdy AP, Mottola MF, Skow RJ, Meah VL, Poitras VJ, et al. Impact of prenatal exercise on both prenatal and postnatal anxiety and depressive symptoms: a systematic review and meta-analysis. Br J Sports Med. 2018;52:1376–1385.
210. Cortes M, Cao M, Liu HL, Moore CS, Durosier LD, Burns P, et al. α7 nicotinic acetylcholine receptor signaling modulates the inflammatory phenotype of fetal brain microglia: first evidence of interference by iron homeostasis. Sci Rep. 2017;7:10645.
211. Frasch MG, Burns P, Benito J, Cortes M, Cao M, Fecteau G, et al. Sculpting the sculptors: methods for studying the fetal cholinergic signaling on systems and cellular scales. Methods Mol Biol Clifton NJ. 2018;1781:341–352.
212. Roubinov DS, Luecken LJ, Curci SG, Somers JA, Winstone LK. A prenatal programming perspective on the intergenerational transmission of maternal adverse childhood experiences to offspring health problems. Am Psychol. 2021;76:337–349.
213. Colizzi M, Lasalvia A, Ruggeri M. Prevention and early intervention in youth mental health: is it time for a multidisciplinary and trans-diagnostic model for care? Int J Ment Health Syst. 2020;14:23.
214. Izett E, Rooney R, Prescott SL, De Palma M, McDevitt M. Prevention of mental health difficulties for children aged 0–3 years: a review. Front Psychol. 2021;11:500361.
215. Al Sager A, Goodman SH, Jeong J, Bain PA, Ahun MN. Effects of multi-component parenting and parental mental health interventions on early childhood development and parent outcomes: a systematic review and meta-analysis. Lancet Child Adolesc Health. 2024;8:656–669.
216. Hirve R, Adams C, Kelly CB, McAullay D, Hurt L, Edmond KM, et al. Effect of early childhood development interventions delivered by healthcare providers to improve cognitive outcomes in children at 0–36 months: a systematic review and meta-analysis. Arch Dis Child. 2023;108:247–257.
217. Ravindran N, Zhang X, Ku S. Within-person bidirectional associations between maternal cortisol reactivity and harsh parenting across infancy and toddlerhood. J Fam Psychol. 2024;38:911–920.
218. Romanowicz M, Saliba MT, Wilton AR, Garzon Hincapie JF, Croarkin KS, Saliba CT, et al. Feasibility of digital augmentation of parent-child interaction therapy: a randomized clinical trial. JAMA Netw Open. 2025;8:e2548869.
219. Wakschlag LS, Roberts MY, Flynn RM, Smith JD, Krogh-Jespersen S, Kaat AJ, et al. Future directions for early childhood prevention of mental disorders: a road map to mental health, earlier. J Clin Child Adolesc Psychol. 2019;48:539–554.
220. Shephard E, Zuccolo PF, Idrees I, Godoy PBG, Salomone E, Ferrante C, et al. Systematic review and meta-analysis: the science of early-life precursors and interventions for attention-deficit/hyperactivity disorder. J Am Acad Child Adolesc Psychiatry. 2022;61:187–226.

221. Hudson JL, Minihan S, Chen W, Carl T, Fu M, Tully L, et al. Interventions for young children's mental health: a review of reviews. Clin Child Fam Psychol Rev. 2023;26:593–641.
222. Frasch MG, Yoon B-J, Helbing DL, Snir G, Antonelli MC, Bauer R. Autism spectrum disorder: a neuro-immunometabolic hypothesis of the developmental origins. Biology. 2023;12:914.
223. Salazar De Pablo G, De Micheli A, Nieman DH, Correll CU, Kessing LV, Pfennig A, et al. Universal and selective interventions to promote good mental health in young people: systematic review and meta-analysis. Eur Neuropsychopharmacol. 2020;41:28–39.
224. Gerstl B, Ahinkorah BO, Nguyen TP, John JR, Hawker P, Winata T, et al. Evidence-based long term interventions targeting acute mental health presentations for children and adolescents: systematic review. Front Psychiatry. 2024;15:1324220.
225. Parlatini V, Bellato A, Gabellone A, Margari L, Marzulli L, Matera E, et al. A state-of-the-art overview of candidate diagnostic biomarkers for attention-deficit/hyperactivity disorder (ADHD). Expert Rev Mol Diagn. 2024;24:259–271.
226. Stolfi F, Abreu H, Sinella R, Nembrini S, Centonze S, Landra V, et al. Omics approaches open new horizons in major depressive disorder: from biomarkers to precision medicine. Front Psychiatry. 2024;15:1422939.
227. Mokhtari A, Ibrahim EC, Gloaguen A, Barrot CC, Cohen D, Derouin M, et al. Using multiomic integration to improve blood biomarkers of major depressive disorder: a case-control study. EBioMedicine. 2025;113:105569.
228. Miller JG, Armstrong-Carter E, Balter L, Lorah J. A meta-analysis of mother-child synchrony in respiratory sinus arrhythmia and contextual risk. Dev Psychobiol. 2023;65:e22355.
229. Oshri A, Liu S, Suveg CM, Caughy MO, Goodgame Huffman L. Biological sensitivity to context as a dyadic construct: An investigation of child–parent RSA synchrony among low-SES youth. Dev Psychopathol. 2023;35:95–108.
230. Butterfield RD, Silk JS, Lee KH, Siegle GS, Dahl RE, Forbes EE, et al. Parents still matter! parental warmth predicts adolescent brain function and anxiety and depressive symptoms 2 years later. Dev Psychopathol. 2020;33:226–239.
231. Jeong J, Franchett EE, Ramos de Oliveira CV, Rehmani K, Yousafzai AK. Parenting interventions to promote early child development in the first three years of life: a global systematic review and meta-analysis. Plos Med. 2021;18:e1003602.
232. Li J, Qian K, Liu X. The impact and mechanisms of parent-child relationship quality and its changes on adolescent depression: a four-wave longitudinal study. Appl Psychol Health Well Being. 2025;17:e12606.
233. Russotti J, Swerbenski H, Handley ED, Michl-Petzing LC, Cicchetti D, Toth SL. Intergenerational effects of maternal depression and co-occurring antisocial behaviors: the mediating role of parenting-related processes. J Fam Psychol. 2023;37:408–419.
234. Schechter JC, Brennan PA, Smith AK, Stowe ZN, Newport DJ, Johnson KC. Maternal prenatal psychological distress and preschool cognitive functioning: the protective role of positive parental engagement. J Abnorm Child Psychol. 2017;45:249–260.
235. Silvey C, Demir-Lira ÖE, Goldin-Meadow S, Raudenbush SW. Effects of time-varying parent input on children's language outcomes differ for vocabulary and syntax. Psychol Sci. 2021;32:536–548.
236. Tang F, Tracy M, Radigan M, Vásquez E. Trajectories of maternal parenting stress and adolescent behavioral symptoms in unmarried families: the role of family immigration status. J Affect Disord. 2024;367:297–306.
237. Zhu Y, Zhang G, Zhan S. Association of parental adverse childhood experiences with offspring sleep problems: the role of psychological distress and harsh discipline. Child Adolesc Psychiatry Ment Health. 2024;18:112–11.
238. Birk SL, Stewart L, Olino TM. Parent–child synchrony after early childhood: a systematic review. Clin Child Fam Psychol Rev. 2022;25:529–551.
239. Wass S, Greenwood E, Esposito G, Smith C, Necef I, Phillips E. Annual research review: 'there, the dance is - at the still point of the turning world' - dynamic systems perspectives on coregulation and dysregulation during early development. J Child Psychol Psychiatry. 2024;65:481–507.
240. Schoppe-Sullivan SJ, Shafer K, Olofson EL, Kamp Dush CM. Fathers' parenting and coparenting behavior in dual-earner families: contributions of traditional masculinity, father nurturing role beliefs, and maternal gate closing. Psychol Men Masculinity. 2021;22:538–550.
241. Cooke JE, Racine N, Plamondon A, Tough S, Madigan S. Maternal adverse childhood experiences, attachment style, and mental health: pathways of transmission to child behavior problems. Child Abuse Negl. 2019;93:27–37.
242. Racine N, Plamondon A, Madigan S, McDonald S, Tough S. Maternal adverse childhood experiences and infant development. Pediatrics. 2018;141:e20172495.
243. Madigan S, Wade M, Plamondon A, Maguire JL, Jenkins JM. Maternal adverse childhood experience and infant health: biomedical and psychosocial risks as intermediary mechanisms. J Pediatr. 2017;187:282–289.e1.
244. Antonelli MC, Frasch MG, Rumi M, Sharma R, Zimmermann P, Molinet MS, et al. Early biomarkers and intervention programs for the infant exposed to prenatal stress. Curr Neuropharmacol. 2022;20:94–106.
245. Álvarez-Mejía D, Rodas JA, Leon-Rojas JE. From womb to mind: prenatal epigenetic influences on mental health disorders. Int J Mol Sci. 2025;26:6096.
246. Zhao N, Tang M, Wang L, Liu T, Zhao T, Xie K, et al. Effects of childhood trauma on depression and cognitive function in first-diagnosed, drug-naïve depressed patients: an observational case-control study. BMC Psychiatry. 2025;25:329.
247. Luby JL, Rogers C, McLaughlin KA. Environmental conditions to promote healthy childhood brain/behavioral development: informing early preventive interventions for delivery in routine care. Biol Psychiatry Glob Open Sci. 2022;2:233–241.

248. Cui Z, Duprey EB, Huffman LG, Liu S, Smith EP, Caughy MO, et al. Neighborhood socioeconomic disadvantage and physical disorder, parenting strategies, and youths' future orientation. J Appl Dev Psychol. 2025;96:101730.
249. Kassa GM, Yu Z, Minwuyelet F, Gross D. Behavioural interventions targeting the prevention and treatment of young children's mental health problems in low- and middle-income countries: a scoping review. J Glob Health. 2025;15:04018.
250. DePasquale CE, Herzberg MP, Gunnar MR. The pubertal stress recalibration hypothesis: potential neural and behavioral consequences. Child Dev Perspect. 2021;15:249–256.
251. Perry NB, Donzella B, Gunnar MR. Pubertal stress recalibration and later social and emotional adjustment among adolescents: the role of early life stress. Psychoneuroendocrinology. 2022;135:105578.
252. Duda JM, Keding TJ, Kribakaran S, Odriozola P, Kitt ER, Cohodes EM, et al. Exposure to unpredictable childhood environments is associated with amygdala activation during early extinction in adulthood. Dev Cogn Neurosci. 2025;74:101578.
253. Tottenham N. Neural meaning making, prediction, and prefrontal-subcortical development following early adverse caregiving. Dev Psychopathol. 2020;32:1563–1578.
254. Farina B, Schimmenti A. The psychopathological domains of attachment trauma: a proposal for a clinical conceptualization. Clin Neuropsychiatry. 2025;22:351–373.
255. Nelson CA, Sullivan EF, Valdes V. Early adversity alters brain architecture and increases susceptibility to mental health disorders. Nat Rev Neurosci. 2025;26:642–656.
256. Yu J, Haynie DL, Gilman SE. Patterns of adverse childhood experiences and neurocognitive development. JAMA Pediatr. 2024;178:678–687.
257. Luby JL, Barch D, Whalen D, Tillman R, Belden A. Association between early life adversity and risk for poor emotional and physical health in adolescence: a putative mechanistic neurodevelopmental pathway. JAMA Pediatr. 2017;171:1168–1175.
258. Nweze T, Ezenwa M, Ajaelu C, Okoye C. Childhood mental health difficulties mediate the long-term association between early-life adversity at age 3 and poorer cognitive functioning at ages 11 and 14. J Child Psychol Psychiatry. 2023;64:952–965.
259. Luby JL, Baram TZ, Rogers CE, Barch DM. Neurodevelopmental optimization after early-life adversity: cross-species studies to elucidate sensitive periods and brain mechanisms to inform early intervention. Trends Neurosci. 2020;43:744–751.
260. Gee DG. Neurodevelopmental mechanisms linking early experiences and mental health: translating science to promote well-being among youth. Am Psychol. 2022;77:1033–1045.
261. Medina-Martínez J, Saus-Ortega C, Sánchez-Lorente MM, Sosa-Palanca EM, García-Martínez P, Mármol-López MI. Health inequities in LGBT people and nursing interventions to reduce them: a systematic review. Int J Environ Res Public Health. 2021;18:11801.
262. Carvalho Silva R, Pisanu C, Maffioletti E, Menesello V, Bortolomasi M. PROMPT consortium *et al*. Biological markers of sex-based differences in major depressive disorder and in antidepressant response. Eur Neuropsychopharmacol J Eur Coll Neuropsychopharmacol. 2023;76:89–107.
263. Ponton E, Turecki G, Nagy C. Sex differences in the behavioral, molecular, and structural effects of ketamine treatment in depression. Int J Neuropsychopharmacol. 2022;25:75–84.
264. Peters A, McEwen BS, Friston K. Uncertainty and stress: why it causes diseases and how it is mastered by the brain. Prog Neurobiol. 2017;156:164–188.
265. Potts C, Kealy C, McNulty JM, Madrid-Cagigal A, Wilson T, Mulvenna MD, et al. Digital mental health interventions for young people aged 16-25 years: scoping review. J Med Internet Res. 2025;27:e72892.
266. Uhlhaas PJ, Davey CG, Mehta UM, Shah J, Torous J, Allen NB, et al. Towards a youth mental health paradigm: a perspective and roadmap. Mol Psychiatry. 2023;28:3171–3181.
267. Uhlhaas PJ, Bechdolf A, Konrad K, Koutsouleris N, Meisenzahl E, Meyer-Lindenberg A, et al. "Youth Mental Health": Früherkennung und Frühintervention psychischer erkrankungen im Jugendalter. Nervenarzt. 2026;97:278–284.
268. McGorry PD, Mei C. Early intervention in youth mental health: progress and future directions. Evid Based Ment Health. 2018;21:182–184.
269. Venema M, Klaassen D, Kwee J, Rossen L. Grieving in community: accompanying bereaved parents. J Community Psychol. 2023;51:2246–2260.
270. Baysarowich R, Humes R, Goez H, Remedios J, Denomey N, DeCoste S, et al. Socioeconomic status and brain development: insights and theoretical perspectives on deficit, adaptation, and resilience. Curr Opin Behav Sci. 2025;63:101502.
271. Theron L, Bergamini M, Chambers C, Choi K, Fawole OI, Fyneface FD, et al. Multisystemic resilience and its impact on youth mental health: reflections on co-designing a multi-disciplinary, participatory study. Front Child Adolesc Psychiatry. 2025;4:1489950.
272. Desplats P, Gutierrez AM, Antonelli MC, Frasch MG. Microglial memory of early life stress and inflammation: susceptibility to neurodegeneration in adulthood. Neurosci Biobehav Rev. 2020;117:232–242.
273. Schafer DP, Lehrman EK, Kautzman AG, Koyama R, Mardinly AR, Yamasaki R, et al. Microglia sculpt postnatal neural circuits in an activity and complement-dependent manner. Neuron. 2012;74:691–705.
274. McCormack K, Bramlett S, Morin EL, Siebert ER, Guzman D, Howell B, et al. Long-term effects of adverse maternal care on hypothalamic-pituitary-adrenal (HPA) axis function of juvenile and adolescent macaques. Biology. 2025;14:204.

275. Meaney MJ. Epigenetics and the biological definition of gene X environment interactions. Child Dev. 2010;81:41–79.
276. Braithwaite E, Kundakovic M, Ramchandani P, Murphy S, Champagne F. Maternal prenatal depressive symptoms predict infant NR3C1 1F and BDNF IV DNA methylation. Epigenetics. 2015;10:408–417.
277. Hollins SL, Cairns MJ. MicroRNA: Small RNA mediators of the brains genomic response to environmental stress. Prog Neurobiol. 2016;143:61–81.
278. Bock J Perinatal epigenetic programming of functional brain circuits. In: Epigenetics in Biological Communication. 2024, pp. 197–218.
279. Bearer EL. Epigenetic changes, a biomarker for non-accidental injury. Pediatr Res. 2023;94:17–19.
280. Bearer EL, Mulligan BS. Epigenetic changes associated with early life experiences: saliva, a biospecimen for DNA methylation signatures. Curr Genomics. 2018;19:676–698.
281. Babenko O, Golubov A, Ilnytskyy Y, Kovalchuk I, Metz GA. Genomic and epigenomic responses to chronic stress involve miRNA-mediated programming. PloS One. 2012;7:e29441.
282. Ambeskovic M, Ilnytskyy Y, Kiss D, Currie C, Montina T, Kovalchuk I, et al. Ancestral stress programs sex-specific biological aging trajectories and non-communicable disease risk. Aging. 2020;12:3828–3847.
283. Audet M-C, McQuaid RJ, Merali Z, Anisman H. Cytokine variations and mood disorders: influence of social stressors and social support. Front Neurosci. 2014;8:416.
284. Niraula A, Witcher KG, Sheridan JF, Godbout JP. Interleukin-6 induced by social stress promotes a unique transcriptional signature in the monocytes that facilitate anxiety. Biol Psychiatry. 2019;85:679–689.
285. Bailey MT, Kinsey SG, Padgett DA, Sheridan JF, Leblebicioglu B. Social stress enhances IL-1beta and TNF-alpha production by porphyromonas gingivalis lipopolysaccharide-stimulated CD11b+ cells. Physiol Behav. 2009;98:351–358.
286. Xu J-J, Guo S, Xue R, Xiao L, Kou J-N, Liu Y-Q, et al. Adalimumab ameliorates memory impairments and neuroinflammation in chronic cerebral hypoperfusion rats. Aging. 2021;13:14001–14014.
287. Skaper SD, Facci L, Zusso M, Giusti P. Neuroinflammation, mast cells, and glia: dangerous liaisons. Neurosci Rev J Bringing Neurobiol Neurol Psychiatry. 2017;23:478–498.
288. Sălcudean A, Bodo C-R, Popovici R-A, Cozma M-M, Păcurar M, Crăciun R-E, et al. Neuroinflammation-A crucial factor in the pathophysiology of depression-A comprehensive review. Biomolecules. 2025;15:502.
289. Bürger Z, Müller VI, Hoffstaedter F, Habel U, Gur RC, Windischberger C, et al. Stressor-specific sex differences in amygdala-frontal cortex networks. J Clin Med. 2023;12:865.
290. Migicovsky Z, Kovalchuk I. Epigenetic memory in mammals. Front Genet. 2011;2:28.
291. Nie C, Sun Y, Zhen H, Guo M, Ye J, Liu Z, et al. Differential expression of plasma Exo-miRNA in neurodegenerative diseases by next-generation sequencing. Front Neurosci. 2020;14:438.
292. Beger RD, Dunn W, Schmidt MA, Gross SS, Kirwan JA, Cascante M, et al. Metabolomics enables precision medicine: 'A White Paper, Community Perspective'. Metabolomics Off J Metabolomic Soc. 2016;12:149.
293. Metz GAS, Ng JWY, Kovalchuk I, Olson DM. Ancestral experience as a game changer in stress vulnerability and disease outcomes. BioEssays News Rev Mol Cell Dev Biol. 2015;37:602–611.
294. Derijk RH, de Kloet ER. Corticosteroid receptor polymorphisms: determinants of vulnerability and resilience. Eur J Pharmacol. 2008;583:303–311.
295. Faraji J, Soltanpour N, Lotfi H, Moeeini R, Moharreri A-R, Roudaki S, et al. Lack of social support raises stress vulnerability in rats with a history of ancestral stress. Sci Rep. 2017;7:5277.
296. Faraji J, Karimi M, Soltanpour N, Moharrerie A, Rouhzadeh Z, Lotfi H, et al. Oxytocin-mediated social enrichment promotes longer telomeres and novelty seeking. eLife. 2018;7:e40262.
297. Sharma D, Bhave S, Gregg E, Uht R. Dexamethasone induces a putative repressor complex and chromatin modifications in the CRH promoter. Mol Endocrinol. 2013;27:1142–1152.
298. Faraji J, Karimi M, Soltanpour N, Rouhzadeh Z, Roudaki S, Hosseini SA, et al. Intergenerational sex-specific transmission of maternal social experience. Sci Rep. 2018;8:10529.

**Publisher's note** Springer Nature remains neutral with regard to jurisdictional claims in published maps and institutional affiliations.

## Authors and Affiliations

**Mariela Chertoff[1,2] · Martin G. Frasch[3,4] · Eduardo T. Cánepa[1,2] · Gerlinde A. S. Metz[5] · Marta C. Antonelli[6] · Sheehan D. Fisher[7] · Bea R. H. Van den Bergh[8]**

[1] Laboratorio de Neuroepigenética y Adversidades Tempranas, Departamento de Química Biológica, Facultad de Ciencias Exactas y Naturales, Universidad de Buenos Aires, Buenos Aires, Argentina

[2] Instituto de Química Biológica de la Facultad de Ciencias Exactas y Naturales (IQUIBICEN), CONICET-Universidad de

Buenos Aires, Facultad de Ciencias Exactas y Naturales, Buenos Aires, Argentina

[3] Department of Obstetrics and Gynecology and Institute on Human Development and Disability, University of Washington, Seattle, WA, USA

[4] Health Stream Analytics, LLC, Seattle, WA, USA

[5] Department of Neuroscience and Canadian Centre for Behavioural Neuroscience, University of Lethbridge, Lethbridge, Canada

[6] Laboratorio de Programación Perinatal del Neurodesarrollo. Instituto de Biología Celular y Neurociencias “Prof. E. De Robertis”. Facultad de Medicina. Universidad de Buenos Aires, Buenos Aires, Argentina

[7] Department of Psychiatry & Behavioral Sciences, Feinberg School of Medicine, Northwestern University, Chicago, USA

[8] Research Group Health Psychology and Leuven Brain Institute, Faculty of Psychology and Educational Science, Leuven, Belgium